\documentclass[10pt,journal]{IEEEtran}
\def\BibTeX{{\rm B\kern-.05em{\sc i\kern-.025em b}\kern-.08em T\kern-.1667em\lower.7ex\hbox{E}\kern-.125emX}}

\usepackage{lipsum}
\usepackage{url}
\usepackage{xcolor,cite,etoolbox}
\ifCLASSINFOpdf
  \usepackage[pdftex]{graphicx}
  \graphicspath{{../pdf/}{../jpeg/}}
  \DeclareGraphicsExtensions{.pdf,.jpeg,.png}
\else
  \usepackage[dvips]{graphicx}
  \graphicspath{{../eps/}}
  \DeclareGraphicsExtensions{.eps}
\fi

\ifCLASSOPTIONcompsoc
\usepackage[caption=false,font=normalsize,labelfont=sf,textfont=sf,subrefformat=parens,labelformat=parens]{subfig}
\else
\usepackage[caption=false,font=footnotesize,subrefformat=parens,labelformat=parens]{subfig}
\fi
\usepackage{bm}
\usepackage{bbm}
\usepackage{dsfont}
\usepackage{amsmath}
\usepackage{amssymb}
\usepackage[thinc]{esdiff}
\usepackage{mathtools}
\usepackage{amsmath,amsfonts,amssymb}
\usepackage{amsmath,amssymb,amsthm}
\DeclareMathOperator{\rank}{rank}

\DeclarePairedDelimiterX{\expectarg}[1]{[}{]}{%
  \ifnum\currentgrouptype=16 \else\begingroup\fi
  \activatebar#1
  \ifnum\currentgrouptype=16 \else\endgroup\fi
}

\newcommand{\innermid}{\nonscript\;\delimsize\vert\nonscript\;}
\newcommand{\activatebar}{%
  \begingroup\lccode`\~=`\|
  \lowercase{\endgroup\let~}\innermid
  \mathcode`|=\string"8000
}

\DeclarePairedDelimiter\norm{\lVert}{\rVert}%
\DeclareMathOperator{\Tr}{Tr}

\usepackage{algorithmic}
\usepackage{algorithm}
\usepackage{url}
\usepackage{epstopdf}
\usepackage{amsfonts}
\usepackage{enumitem}
\usepackage{array}
\usepackage{xcolor, soul}
\newcolumntype{L}[1]{>{\raggedright\let\newline\\\arraybackslash\hspace{0pt}}m{#1}}
\newcolumntype{C}[1]{>{\centering\let\newline\\\arraybackslash\hspace{0pt}}m{#1}}
\newcolumntype{R}[1]{>{\raggedleft\let\newline\\\arraybackslash\hspace{0pt}}m{#1}}
\begin{document}

\title{Rate-Splitting for Intelligent Reflecting Surface-Aided Multiuser VR Streaming}

\author{Rui Huang, \textit{Student Member, IEEE}, Vincent W.S. Wong, \textit{Fellow, IEEE}, and  Robert Schober, \textit{Fellow, IEEE}\\

\thanks{
R. Huang and V. W.S. Wong are with the Department of Electrical and Computer Engineering, The University of British Columbia, Vancouver, BC, V6T 1Z4, Canada (e-mail: \{ruihuang, vincentw\}@ece.ubc.ca). 

R. Schober is with the Institute for Digital Communications, Friedrich-Alexander University of Erlangen-Nuremberg, Erlangen 91058, Germany (e-mail: robert.schober@fau.de). 

   }
}

\maketitle
\vspace{-15mm}

\IEEEpeerreviewmaketitle
\newcommand{\C}{\mathbb{C}}
\newcommand{\D}{\mathcal{D}}
\newcommand{\Sta}{\boldsymbol{S}}
\newcommand{\Sty}{\boldsymbol{Y}}
\newcommand{\Stz}{\boldsymbol{z}}
\newcommand{\R}{\mathcal{R}}
\newcommand{\M}{\mathcal{M}}
\newcommand{\W}{\mathcal{W}}
\newcommand{\bias}{\boldsymbol{b}}
\newcommand{\K}{\mathcal{K}}
\newcommand{\uset}{\mathcal{U}}
\newcommand{\Los}{\mathcal{L}}
\newcommand{\T}{\mathcal{T}}
\newcommand{\N}{\mathcal{N}}
\newcommand{\I}{\mathcal{I}}
\newcommand{\V}{\mathcal{V}}
\newcommand{\F}{\boldsymbol{G}}
\newcommand{\Q}{Q}
\newcommand{\que}{\boldsymbol{q}(t)}
\newcommand{\n}{\boldsymbol{n}}
\newcommand{\TA}{\boldsymbol{\Phi}}
\newcommand{\MP}{\boldsymbol{\Psi}}
\newcommand{\TAO}{\boldsymbol{\widehat{\Phi}}}
\newcommand{\PS}{\boldsymbol{\psi}}
\newcommand{\THRES}{\Gamma^{\text{min}}}
\newcommand{\Pvec}{\boldsymbol{P}^{\text{tx}}}
\newcommand{\ptx}{P^{\text{tx}}}
\newcommand{\G}{\mathcal{G}}
\newcommand{\br}{\boldsymbol{v}}
\newcommand{\lf}{\mathcal{L}}
\newcommand{\E}{\mathcal{E}}
\newcommand{\pmax}{P^{\text{max}}}
\newcommand{\aH}{\boldsymbol{H}}
\graphicspath{{figure/}}

\newcommand{\pset}{\mathcal{L}}
\newcommand{\lossr}[1]{\ell(r_#1(t))}
\newcommand{\h}{\boldsymbol{h}}
\newcommand{\psm}{\boldsymbol{\Psi}}
\newcommand{\pol}{\boldsymbol{p}}
\newcommand{\psmv}{\boldsymbol{\psi}}
\newcommand{\psmr}{\boldsymbol{\phi}}
\newcommand{\diag}{\text{diag}}
\newcommand{\rhob}{\rho_\mathrm{L}(t)}
\newcommand{\X}{\boldsymbol{x}(t)}
\newcommand{\Y}{\boldsymbol{y}(t)}
\newcommand{\Z}{\boldsymbol{z}(t)}
\newcommand{\ut}{\boldsymbol{u}}
\newcommand{\bfv}{\boldsymbol{b}}
\newcommand{\rbfv}{\boldsymbol{m}}
\newcommand{\rbfm}{\boldsymbol{M}}
\newcommand{\controlstatespace}{\mathcal{H}}
\newcommand{\PI}{\boldsymbol{\Pi}(t)}
\newcommand{\PSIfunc}[1]{\hat{\Psi}_{#1}}
\newcommand{\nexp}[1]{e^{#1}}
\newcommand{\rhoal}{\rho_\mathrm{H}(t)}
\newcommand{\usv}{\boldsymbol{x}}
\newcommand{\w}{\boldsymbol{w}}
\newcommand{\s}{\boldsymbol{s}}
\newcommand{\Gm}{\boldsymbol{G}}
\newcommand{\B}{\boldsymbol{b}}
\newcommand{\cv}{\boldsymbol{c}}
\newcommand{\TD}{T_{\text{DL}}}
\newcommand{\ctxt}{\text{c}}
\newcommand{\ptxt}{\text{p}}
\newcommand{\sta}{\boldsymbol{s}}
\newcommand{\act}{\boldsymbol{a}}

\newcommand{\vr}{\boldsymbol{v}}

\newcommand{\pgs}{\boldsymbol{s}^{\text{PB}}}
\newcommand{\pga}{\boldsymbol{a}^{\text{PB}}}
\newcommand{\pgr}{r^{\text{PB}}}
\begin{abstract}
The growing demand for virtual reality (VR) applications requires wireless systems to provide a high transmission rate to support 360-degree video streaming to multiple users simultaneously.
In this paper, we propose an intelligent reflecting surface (IRS)-aided rate-splitting (RS) VR streaming system.
In the proposed system, RS facilitates the exploitation of the shared interests of the users in VR streaming, and IRS creates additional propagation channels to support the transmission of high-resolution 360-degree videos.
IRS also enhances the capability to mitigate the performance bottleneck caused by the requirement that all RS users have to be able to decode the common message.
We formulate an optimization problem for maximization of the achievable bitrate of the 360-degree video subject to the quality-of-service (QoS) constraints of the users.
We propose a deep deterministic policy gradient with imitation learning (Deep-GRAIL) algorithm, in which we leverage deep reinforcement learning (DRL) and the hidden convexity of the formulated problem to optimize the IRS phase shifts, RS parameters, beamforming vectors, and bitrate selection of the 360-degree video tiles.
We also propose RavNet, which is a deep neural network customized for the policy learning in our Deep-GRAIL algorithm.
Performance evaluation based on a real-world VR streaming dataset shows that the proposed IRS-aided RS VR streaming system outperforms several baseline schemes in terms of system sum-rate, achievable bitrate of the 360-degree videos, and online execution runtime.
Our results also reveal the respective performance gains obtained from RS and IRS for improving the QoS in multiuser VR streaming systems.

\end{abstract}
%\newpage
\section{Introduction}
Virtual reality (VR) streaming provides the users with an immersive experience by rendering 360-degree videos using head-mounted devices (HMDs).
VR is considered as one of the important use cases of the sixth generation (6G) of wireless systems \cite{walid2020vision}.
Via the wireless connectivity, VR users can move and interact freely without being restricted by the cable that connects the HMDs and VR server.
Driven by the development of VR technology, there are emerging applications of VR streaming in different industries, including entertainment, retail, and education.
It is estimated that the global VR market size will increase from \$16.67 billion US dollars (USD) in 2022 to \$227.34 billion USD by 2029, with a compound annual growth rate of 45.2\% \cite{marketsize_art}.

The increasing number of VR users and the growing demand for VR streaming introduce new challenges to the current wireless systems.
First, the bitrate of a high-resolution 360-degree video can be much higher than that of conventional multimedia applications.
For example, a 4K 360-degree video may have a bitrate of 78 mega bits per second (Mbps) \cite{goprobitrate_art}.
In addition, a VR user may experience motion sickness when the motion-to-photon delay, i.e., the delay between the head movement and the requested 360-degree video segments being rendered at the HMD of this user, is larger than 20 ms \cite{elbamby2018toward}.
To mitigate these issues, the data transmission of 360-degree videos should be completed within a short downlink transmission window.
Hence, wireless systems have to be able to support a high data transmission rate to meet the requirement of 360-degree video streaming.

%In this paper, we tackle the aforementioned challenges by using rate-splitting (RS), intelligent reflecting surface (IRS), and deep reinforcement learning (DRL) techniques, and exploiting the shared interests of the users in a multiuser VR streaming system.
In multiuser VR streaming, the same 360-degree video segment may be requested by multiple users due to their shared interests.
As an example, for the streaming of a 360-degree soccer match video, the supporters of a particular soccer team may frequently share those field-of-views (FoVs) that include the players of their team.
Thus, the data requested by the users are \textit{correlated} due to the shared interests of the users in the same video tiles in multiuser VR streaming systems.
In this paper, we use rate-splitting (RS) to take advantage of the correlations resulting from the shared interests of the users, with the objective to achieve an additional multiplexing gain to improve the VR streaming quality.
RS is a physical layer technique in which the information intended for the users is split into two parts, namely a common message and private messages \cite{clerckx2021open,hamdi2016sum,mao2021rate}.
In RS, each user needs to decode the common message first.
The user then subtracts the common message from the received signal using successive interference cancellation (SIC) and subsequently decodes its private message \cite{hamdi2016sum}.
These features of RS make it a promising technique for multiuser VR streaming systems since (a) the data related to the shared interests of the users can be encoded into the common message to exploit the correlations between the data requested by the users, and (b) the unique data requested by each user can be encoded in its private message to ensure that all the requested video tiles can be received by the user.
% \textcolor{blue}{
% While other physical layer techniques, such as wireless relaying \cite{Kang2022irs}, can also be applied to exploit the shared interests, they may incur additional rounds of data encoding, transmission, and decoding during VR streaming, which can introduce extra delay and computational overhead.}
In this paper, we show that the quality-of-service (QoS) in a multiuser VR streaming system can be significantly improved with a properly designed RS scheme that facilitates the exploitation of the shared interests of the users.

In multiuser VR streaming systems, a low signal-to-interference-plus-noise ratio (SINR) may be experienced by  users who are located far from the base station, resulting in a large path loss and a weak channel gain, as well as users who are located close to other users, thereby experiencing significant interference.
A low SINR can lead to a significant performance degradation in RS systems since the transmission rate of the common message is limited by the user experiencing the minimum SINR \cite{clerckx2021open,hamdi2016sum,mao2021rate}.
Therefore, if the SINR of one user is low, only a small amount of data can be transmitted via the common message to exploit the shared interests of the users, and therefore only a limited multiplexing gain can be achieved by using RS.

To address this issue, we propose to deploy an intelligent reflecting surface (IRS) to increase the minimum SINR of the users and thereby improving the performance of the RS system.
IRSs are reconfigurable planar surfaces with a large number of passive reflecting elements.
Each reflecting element can apply an independent phase shift to the incident signal to reflect the phase-shifted signal towards the receiver.
IRSs can effectively improve the minimum SINR in RS systems because   users who suffer large path loss can benefit from the additional propagation channels (i.e., reflected channels) created by the IRSs.
IRSs also introduce additional degrees of freedom (DoF) (i.e., the phase shifts of the reflecting elements) that can be exploited to mitigate interference \cite{wu2019irs}.
Furthermore, IRSs can be installed on the walls of indoor facilities for VR streaming, making them flexible and efficient extensions for existing VR streaming systems.

In this paper, we propose an IRS-aided RS VR streaming system, where RS is applied to exploit the shared VR streaming interests of the users, and IRSs are used to improve the minimum SINR experienced by the common message across the users and the system sum-rate.
We aim to maximize the achievable bitrate of the 360-degree video by optimizing the IRS phase shifts, RS parameters, beamforming vectors, and individual bitrates.
Solving such a problem using conventional optimization methods (e.g., alternating optimization (AO)) can be computationally expensive and time-consuming for VR streaming systems.
In addition, due to the nonconvexity of the joint optimization problem, some of the optimization variables need to be relaxed.
Such a relaxation (e.g., semidefinite relaxation (SDR)) may incur  performance degradation, rendering the obtained solution to be suboptimal.
To tackle these issues, we propose a deep deterministic policy gradient with imitation learning (Deep-GRAIL) algorithm, in which deep reinforcement learning (DRL) is used to solve the formulated constrained optimization problem in a computationally efficient manner.
Using DRL, the solutions can be obtained based on the forward propagation of the deep neural network (DNN), which in general requires fewer matrix multiplications than conventional optimization methods.
The forward propagation of the DNNs can be accelerated by existing software and hardware design methods \cite{cuda}.
Furthermore, DRL can be applied to solve either convex or nonconvex optimization problems, without necessarily relying on the (hidden) convexity of the problems.
Apart from DRL, we also use imitation learning \cite{rashidinejad2021bridging,nair2018demo} in the proposed Deep-GRAIL algorithm.
Imitation learning allows the learning agent to learn not only from its own exploration, but also from the iterative problem-solving process of the conventional optimization methods.
The contributions of this paper are as follows:
\begin{itemize}
\item We propose an IRS-aided RS VR streaming system, and formulate a joint optimization problem for maximization of the achievable bitrate of the 360-degree video.
Our problem formulation comprises the joint optimization of the beamforming vectors at the base station, IRS phase shifts, RS parameters, and bitrates of the 360-degree video tiles requested by the users.

\item We propose the Deep-GRAIL algorithm, in which imitation learning, actor-critic method, and deep deterministic policy gradient (DDPG) are exploited to learn a policy for solving the formulated problem.
Apart from the experience replay that maintains the exploration history of the learning agent, we introduce a demonstration replay that stores the solutions obtained by conventional optimization methods.
Using imitation learning, the proposed algorithm can effectively improve the learned policy by exploiting both the experience replay and demonstration replay.

\item We propose RavNet, which is a DNN module designed for policy learning in the proposed Deep-GRAIL algorithm.
In particular, one of the neural network layers in RavNet is the differentiable convex optimization (DCO) layer \cite{agrawal2019}, which tackles the convex constraints of the formulated problem during the learning process.

\item We evaluate the performance of the proposed Deep-GRAIL algorithm using a real-world VR streaming dataset \cite{knorr2018vr}.
Simulation results show that the proposed algorithm outperforms several baseline schemes, including the IRS-aided RS non-orthogonal unicast and multicast (RS-NOUM) system using an AO algorithm \cite{mao2019rate}, the conventional IRS-aided multiuser system using an AO algorithm \cite{wu2019irs}, the IRS-aided RS VR streaming system using an AO algorithm, and the IRS-aided RS VR streaming system using a supervised learning algorithm, in terms of the system sum-rate, achievable bitrate, and runtime.
\end{itemize}

The remainder of this paper is organized as follows. 
The related work is discussed in Section II.
The system model and problem formulation for IRS-aided RS VR streaming systems are presented in Section III. 
In Section IV, we develop the Deep-GRAIL algorithm.
In Section V, we introduce the DNN structure and functionality of the proposed RavNet.
Simulation results are presented in Section VI. 
Conclusions are drawn in Section VII.

{\it Notations}: In this paper, we use upper-case and lower-case boldface letters to denote matrices and column vectors, respectively. 
%We use upper-case calligraphic letters to denote sets.
$\mathbb{C}^{M\times N}$ denotes the set of $M \times N$ complex-valued matrices.
$\boldsymbol{A}^T$ and $\boldsymbol{A}^H$ denote the transpose and conjugate transpose of matrix $\boldsymbol{A}$, respectively.
$\mathrm{vec}(\boldsymbol{A})$ returns a vector obtained by stacking the columns of matrix $\boldsymbol{A}$.
$\mathrm{diag}(\boldsymbol{x})$ returns a diagonal matrix where the diagonal elements are given by the elements of vector $\boldsymbol{x}$.
$\Re(\boldsymbol{x})$ and $\Im(\boldsymbol{x})$ return the vectors that include the real and imaginary parts of the complex-valued elements of vector $\boldsymbol{x}$, respectively. 
$\sim$ means ``distributed as".
$\mathbb{E}[\,\cdot\,]$ represents statistical expectation.
$\mathds{1}(\cdot)$ denotes the indicator function, which is equal to $1$ if its argument is true and is equal to $0$ otherwise.
Key notations are summarized in Table \ref{notationtable}.
\begin{table}[t!]
  \begin{center}
    \caption{List of Key Notations}
    \setlength\extrarowheight{-0pt}
    \begin{tabular}{|C{11mm}|C{65mm}|}
  	  \hline
      Variable & Definition\\
      \hline
      $\boldsymbol{a}$ & Action vector\\
      \hline
      $\bfv_0(t)$ & Beamforming vector for the common message in time slot $t$\\
      \hline
      $\bfv_n(t)$ & Beamforming vector for the private message of user $n$ in time slot $t$\\
      \hline
      $c_i(t)$ & Proportion of $R^{\ctxt}(t)$ that is dedicated to the data transmission of video tile $i$ in time slot $t$\\
      \hline
      $\Gm(t)$ & Channel gain between the base station and the IRS  in time slot $t$\\
      \hline
      $\h_{n,D}(t)$ & Channel gain between the base station and user $n$  in time slot $t$\\
      \hline
      $\h_{n,R}(t)$ & Channel gain between the IRS and user $n$  in time slot $t$\\
      \hline
      $I_{\text{max}}$ & Total number of video tiles in each $360$-degree video frame \\
      \hline
      $\I_n(t)$ & Set that collects the indices of video tiles requested by user $n$  in time slot $t$ \\
      \hline
      $I_n(t)$ & Number of video tiles requested by user $n$  in time slot $t$\\
      \hline
      $J_{n,i}(t)$ & Total number of bits that user $n$ received for tile $i$ in time slot $t$\\
      \hline
      $J^{\ctxt}_{n,i}(t)$ & Number of bits that user $n$ obtained from the common message for tile $i$ in time slot $t$\\
      \hline
      $J^{\ptxt}_{n,i}(t)$ & Number of bits that user $n$ obtained from its private message for tile $i$ in time slot $t$\\
      \hline
      $L$ & Number of reflecting elements on the IRS \\
      \hline
      $M$ & Number of available bitrate selections \\
      \hline
      $N$ & Number of users \\
      \hline
      $N_t$ & Number of antennas at the base station \\
      \hline
      $p_{n,i}(t)$ &   Proportion of $R_n^\ptxt(t)$ that is dedicated to the data transmission of video tile $i$ in time slot $t$\\
      \hline
      $r$ & Reward function  \\
      \hline
      $R^{\ctxt}(t)$ & Transmission rate of the common message in time slot $t$\\
      \hline
      $R_n^{\ctxt}(t)$ & Achievable rate for the common message at user $n$ in time slot $t$\\
      \hline
      $R_n^{\ptxt}(t)$ & Achievable rate for the private message at user $n$  in time slot $t$\\
      \hline
      $\boldsymbol{s}$ & State vector \\
      \hline
      $\TD$ & Downlink transmission time duration \\
      \hline
      $T_v$ & Time duration of a video frame \\
      \hline
      $u_n(t)$ & Utility function of user $n$ in time slot $t$\\
      \hline
      $v_{n,i}(t)$ & Bitrate of tile $i$ requested by user $n$  in time slot $t$\\
      \hline
      $\V$ & Set that collects the available bitrate selections\\
      \hline
      $W$ & Bandwidth for downlink transmission \\
      \hline
      $\psm(t)$ & Phase shift control matrix of the IRS  in time slot $t$\\
      \hline
      $\TA_{\text{act}}$ & Learnable parameters of the actor network \\
      \hline
      $\TA^{(m)}_{\text{crt}}$ & Learnable parameters of the $m$-th critic network \\
      \hline
      $\gamma$ & Discount factor \\
      \hline
    \end{tabular}
    \label{notationtable}
  \end{center}
\end{table}

\section{Related Work}
\subsection{Rate Splitting}
Most existing works studied RS systems where the data for different users are independent and uncorrelated \cite{gui2021rate,mao2021rate,ankur2021rate}. 
However, in VR streaming, different users may request the data of the same 360-degree video segment due to their shared interests.
In this case, it becomes important to take the shared interests of the users into account when designing the RS scheme.
However, the shared interests of the users have not been exploited in \cite{gui2021rate,mao2021rate,ankur2021rate}.
The RS VR streaming system we consider in this paper is related to RS multicast systems \cite{hamdi2017rate,mao2019rate}.
The authors in \cite{mao2019rate} considered an RS-NOUM system, where a multicast message needs to be received by all the users in the system and each user's private message is being sent via unicast.
The authors in \cite{hamdi2017rate} studied RS multigroup multicast systems, in which the same message is requested by the users of the same group.
Although the RS schemes considered in \cite{hamdi2017rate,mao2019rate} exploit the multiplexing gain of RS-based multicasting, they assumed that a part of the information is requested by every user in the system (as in \cite{mao2019rate}) or in the same group (as in \cite{hamdi2017rate}).
However, this assumption may not always hold in RS VR streaming systems when the FoVs of some users do not overlap.
Moreover, while it is possible to divide the users into groups based on their FoVs and apply the multigroup RS algorithm in \cite{hamdi2017rate}, finding the optimal user grouping (i.e., determining the number of groups and the number of users within each group) is non-trivial and computationally intensive.
To tackle these issues, the RS parameters in RS VR streaming systems need to be optimized based on the FoVs and the shared interests of the VR users.

\subsection{IRS-aided Wireless Systems}
Existing research has confirmed the benefits of employing IRSs in conventional multiuser wireless communication systems without RS \cite{huang2019ereconf,ma2021joint,huang2022joint,chaccour2020risk,najafi2021physics,besser2022ris}.
The authors in \cite{huang2019ereconf} solved the joint phase shift and power control problem for maximization of the energy efficiency in IRS-aided systems.
Fractional programming (FP) \cite{shen2018fractional} was applied in \cite{ma2021joint} to develop low-complexity beamforming and IRS phase shift algorithms.
The authors in \cite{huang2022joint} proposed a DRL-based algorithm to solve the joint user scheduling, IRS phase shift, and beamforming optimization problem in IRS-aided systems.
The authors in \cite{chaccour2020risk} showed that IRS can improve both the sum-rate and reliability of data transmission in VR applications.
Physics-based modeling of IRS and codebook design for scalable IRS phase shift optimization were studied in \cite{najafi2021physics}.
The authors in \cite{besser2022ris} proposed a phase hopping algorithm for IRS-aided systems to improve the reliability of data transmission without requiring channel state information (CSI).
However, the aforementioned works have not investigated the benefits of combining IRS with RS.
The authors in \cite{ankur2021rate} proposed an IRS-aided rate-splitting multiple access (RSMA) system and designed an on-off control scheme to adjust the phase shift of the IRS.
The authors in \cite{hao2021resource} proposed an AO algorithm to maximize the minimum achievable rate of an IRS-aided multiuser multiple-input single-output (MU-MISO) RSMA system.
However, the application of RS and IRS in multiuser VR streaming systems has not been studied in \cite{ankur2021rate,hao2021resource}.
In IRS-aided RS VR streaming systems, the joint optimization of the IRS phase shifts, RS parameters, beamforming vectors, and bitrate selection of the 360-degree video tiles based on the FoVs and CSI of the VR users is crucial for achieving a high performance.

\subsection{DRL for Multimedia Streaming in Wireless Systems}
DRL has been applied to improve the quality of adaptive bitrate (ABR) video streaming in wireless systems \cite{Wang2022deep,Huang2020learning}.
The authors in \cite{Wang2022deep} studied the ABR video streaming in mobile edge computing (MEC) networks.
They proposed a DRL-based algorithm to jointly optimize the bitrate and transmit power for the videos, as well as the computational resource allocation at the MEC servers.
Moreover, the authors in \cite{Huang2020learning} proposed an algorithm to support ABR video streaming in heterogeneous network conditions.
They trained a meta-model by exploiting meta-reinforcement learning and domain knowledge, such that the meta-model can adapt to specific network conditions after a few training iterations.
Although the results reported in \cite{Wang2022deep,Huang2020learning} show the benefits of using DRL to design bitrate selection algorithms, the DoF offered by the physical layer techniques, i.e., RS and IRS, have not been explored.
In this paper, we demonstrate that the video streaming quality can further be improved by jointly optimizing the DoF of the wireless systems with the bitrate selection of the video tiles.
We also validate the capability of DRL for solving the joint optimization problem in a computationally efficient manner.

\color{black}
\section{IRS-aided RS VR Streaming System and Problem Formulation}
The considered IRS-aided RS VR streaming system is illustrated in Fig. \ref{scenario_just}.
One base station and an IRS are deployed in an indoor facility to provide VR streaming service to $N$ users.
Let $\N=\{1,2,\ldots,N\}$ denote the set of users.
The base station has $N_t$ antennas, while the IRS has $L$ reflecting elements.
The HMD of each user has one antenna.
We denote $\phi_l\in [0, 2\pi),\,l\in\{1,\ldots,L\},$ as the phase shift of the $l$-th reflecting element of the IRS. 
Time is slotted into intervals of equal duration. 
Let $\mathcal{T} = \{1, 2, \ldots\}$ denote the set of time slots. 
The time interval $[t, t + 1)$ is referred to as time slot $t\in\T$.
The direct channel gain between the base station and user $n\in\N$ in time slot $t$ is denoted by $\h_{n,D}(t)\in\mathbb{C}^{N_t}$.
The channel gain between the base station and the IRS in time slot $t$ is denoted by $\Gm(t)\in\mathbb{C}^{L\times N_t}$.
The phase shift matrix of the IRS in time slot $t$ is an $L \times L$ diagonal matrix, denoted by $\psm(t)=\text{diag}(e^{j\phi_1(t)},\ldots,e^{j\phi_{L}(t)})$.
The channel gain between the IRS and user $n\in\N$ in time slot $t$ is denoted as $\h_{n,R}(t)\in\mathbb{C}^{L}$.

To ensure proper functionality of the HMD, each user is designated a certain area (denoted by the yellow areas in Fig. \ref{scenario_just}) inside the indoor facility \cite{oculuslink_art}.
Each user can move freely within his/her designated area during the VR streaming session. 
%We consider an indoor environment where the downlink channels are dominated by the line-of-sight channels between the base station, IRS, and the users.
Since the users can only move within the designated area with a low mobility, we assume the base station can assign orthogonal pilot symbols to the users, and exploit existing channel estimation methods proposed for IRS-aided multiuser systems, such as \cite{wang2020channel,li2021channel,guan2022anchor}, to obtain the global CSI.
In order to investigate the performance upper bound of an IRS-aided RS VR streaming system, we assume that perfect CSI can be obtained by the base station.

\subsection{Video Tile Request of the Users}
\begin{figure}[!t]
\centering
\includegraphics[width=3.45in]{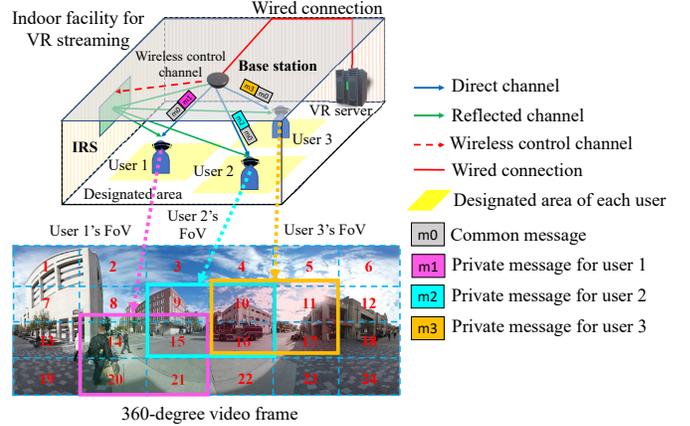}
\caption{An IRS-aided RS VR streaming system. 
The upper part of the figure shows an indoor facility for VR streaming.
The lower part of the figure illustrates a 360-degree video frame.
The $2\times 2$ boxes in the video frame represent the FoVs of the users, while the numbers are the indices of the corresponding 360-degree video tiles. Here, users 1 and 2 request the same video tile with index 15.
Users 2 and 3 request the same video tile with indices 10 and 16.}
\label{scenario_just}
\end{figure}

Each 360-degree video frame is divided into $I_{\text{max}}$ video tiles with $N_x$ rows and $N_y$ columns, i.e., $I_{\text{max}} = N_x N_y$.
As an example, for the 360-degree video frame shown in Fig. \ref{scenario_just}, we have $N_x=4$, $N_y=6$, and $I_{\text{max}}=24$.
We denote $\I=\{1,2,\ldots,I_{\text{max}}\}$ as the set of indices of the tiles.
At the beginning of time slot $t\in\T$, user $n\in\N$ sends an uplink request to the base station that specifies the indices of the tiles it requested, which can be determined based on the FoV of user $n$.
The indices of tiles requested by user $n$ in time slot $t$ are collected in set $\I_n(t)\subseteq \I$.
We denote the number of video tiles requested by user $n$ in time slot $t$ as $I_n(t)$.
We have $|\I_n(t)| = I_n(t)\leq I_{\text{max}}, n\in\N, t\in\T$, where $|\I_n(t)|$ denotes the cardinality of set $\I_n(t)$.

\textit{Remark 1:} 
Although we assume user $n$ informs the base station about $\I_n(t)$ via uplink signaling, the base station can also use existing prediction algorithms to predict $\I_n(t)$ based on historical information, see, e.g., \cite{anh2018your}.
The prediction of video tile requests is beyond the scope of this work but our proposed algorithm can also be straightforwardly applied to VR streaming systems with video tile request prediction.

\subsection{Bitrate Selection, QoS, and User's Utility Model}
The data of the $360$-degree video is stored at the VR server, which is connected to the base station via a wired connection.
We assume there are $M$ bitrate selections of the 360-degree video available at the VR server.
After receiving the video tile requests from the users, the base station needs to determine the bitrate of each requested video tile.
Let $v_{n,i}(t)$ denote the bitrate of tile $i\in\I_n(t)$ requested by user $n$ in time slot $t$.
For bitrate selection, we have
\begin{equation}\label{bitrate_c}
\text{C1:}\quad v_{n,i}(t)\in\V=\{v_1,\ldots,v_{M}\},\,i\in\I_n(t),\,n\in\N,
\end{equation}
where set $\V$ contains the $M$ bitrate selections available at the VR server, and $v_1<v_2<\cdots<v_M$.
We use vector $\br_n(t) = (v_{n,i}(t),\,i\in\I_n(t))$ to collect the bitrate selections of the tiles requested by user $n$ in time slot $t$.

We model the QoS degradation caused by a bitrate switch
among the tiles requested by a user in a particular time slot by the intra-frame quality switch loss \cite{ming2020online,samira2016qoe}.
We denote the intra-frame quality switch loss of user $n$ in time slot $t$ as $\ell^{\text{intra}}_{n}(t)$.
$\ell^{\text{intra}}_{n}(t)$ is determined by the variance of the elements of vector $\br_n(t)$.
For user $n\in\N$ in time slot $t\in\T$, we have \cite{ming2020online}
\begin{equation}
\ell^{\text{intra}}_{n}(t) = \frac{1}{I_n(t)}\sum_{i\in\I_n(t)}\left(v_{n,i}(t)- \frac{1}{I_n(t)}\sum_{j\in\I_n(t)} v_{n,j}(t)\right)^2.
\end{equation}
The utility obtained by user $n$ in time slot $t$ is given by
\begin{equation}\label{user_utility}
u_n(t) = \sum_{i\in\I_n(t)} v_{n,i}(t) - \kappa^{\text{intra}} \ell^{\text{intra}}_{n}(t),\,n\in\N,
\end{equation}
where $\kappa^{\text{intra}}>0$ is a scaling factor for $\ell^{\text{intra}}_{n}(t)$.
The utility in (\ref{user_utility}) captures the QoS improvement obtained with higher bitrates of the requested tiles and the QoS degradation caused by intra-frame quality switches.
The utility function in (\ref{user_utility}) is motivated by the current standardization of video streaming in wireless systems.
Dynamic adaptive streaming over HTTP (DASH) has been standardized by the Third Generation Partnership Project (3GPP) as the protocol for supporting video streaming services in wireless systems \cite{3gpdash}.
The utility function in (\ref{user_utility}) takes into account two important metrics in DASH \cite{dashifprotocol} for measuring  video streaming quality, namely, the achievable bitrate of the video and the bitrate switch during video streaming.
The achievable bitrate of the video is a widely used metric for measuring streaming quality.
The bitrate switch during the video streaming has been recognized as an important factor in video streaming experience based on DASH \cite{Seufert2015survey}.

\subsection{RS-based Downlink VR Tile Transmission}
The base station employs RS-based downlink transmission to exploit the tile requests shared by different users.
In time slot $t$, the indices of the tiles requested by all users are collected in set $\I(t) =\bigcup_{n\in\N} \I_n(t), \,t\in\T$.
After receiving the video tile requests from the users, the base station constructs a common message taking the video tile requests and CSI of the users into account.
When constructing the common message in the proposed IRS-aided RS VR streaming system, the base station needs to determine (a) the data of which tiles should be included in the common message, and (b) what is the proportion of the data of each tile in the common message.
After construction, the common message is encoded into a data stream, which is denoted as $s_0(t)\in\mathbb{C}$, where $\mathbb{E}\left[|s_0(t)|^2\right]=1$.
The beamforming vector for the common message is denoted as $\bfv_0(t)\in\mathbb{C}^{N_t}$.
In addition, the base station constructs a private message for user $n\in\N$ that includes the private part of the data of the tiles requested by user $n$ in time slot $t$. 
The private message for user $n$ is encoded as $s_n(t)\in\mathbb{C}$ with $\mathbb{E}\left[|s_n(t)|^2\right]=1$.
The beamforming vector for the private message of user $n$ is denoted as $\bfv_n(t)\in\mathbb{C}^{N_t}$.
We collect the beamforming vectors in vector $\B(t) = \left[\bfv^T_0(t)\,\,\bfv^T_1(t)\,\,\cdots\,\,\bfv^T_N(t)\right]^T\in\mathbb{C}^{(N+1) N_t}$.
We denote the maximum transmit power of the base station as $P_{\text{max}}$.
We have the following constraint on the beamforming vectors:
\begin{equation}\label{pri_c4}
\text{C2:}\quad ||\bfv(t)||_2^2\leq P_{\text{max}}.
\end{equation}

At the receiver side, the signal received by user $n$ in time slot $t$ is given by:
\begin{equation}
\begin{aligned}
y_n(t) = & \,\,\left(\h^H_{n,D}(t) + \h^H_{n,R}(t)\,\psm(t)\,\Gm(t)\right) \bfv_0(t) s_0(t)  \\
&+ \sum_{m\in\N} \left(\h^H_{n,D}(t)+ \h^H_{n,R}(t)\,\psm(t)\,\Gm(t)\right) \bfv_m(t) s_m(t) \\
&+ z_n(t),\,\,n\in\N,\,\,t\in\T,
\end{aligned}
\end{equation}
where $z_n(t)$ is the additive white Gaussian noise (AWGN) with zero mean and variance $\sigma^2$ at user $n$ in time slot $t$.
User $n$ first decodes the common message by treating the private messages of all users as interference.
The SINR of the common message at user $n$ in time slot $t$ is given by:
\begin{equation}\label{common_sinr}
\begin{aligned}
&\gamma_n^{\ctxt}(t) \\
&= \frac{\big|\left(\h^H_{n,D}(t) + \h^H_{n,R}(t)\,\psm(t)\,\Gm(t)\right)\bfv_0(t)\big|^2}{\sum_{m\in\N} \big|\left(\h^H_{n,D}(t)+ \h^H_{n,R}(t)\,\psm(t)\,\Gm(t)\right)\bfv_m(t) \big|^2 + \sigma^2}.
\end{aligned}
\end{equation}
The achievable rate for the common message at user $n$ in time slot $t$ is $R_n^{\ctxt}(t) = \log_2 (1 + \gamma_n^{\ctxt}(t))$.
Let $R^{\ctxt}(t)$ denote the transmission rate of the common message in time slot $t$.
All users need to decode the common message first, and then remove it from their respective received signal to decode their private message.
%The common message needs to be decoded by all users first, and then be removed from the received signals using SIC such that the private messages can be decoded.
To ensure successful decoding of the common message at all users, we have the following constraint on $R^{\ctxt}(t)$:
\begin{equation}\label{common_rate_c}
\text{C3:}\quad R^{\ctxt}(t) = \min\{R_1^{\ctxt}(t),\ldots,R_N^{\ctxt}(t)\}.
\end{equation}

\noindent We denote the proportion of $R^{\ctxt}(t)$ that is dedicated to the data transmission of video tile $i\in\I(t)$ in time slot $t$ as $c_i(t)$.
We have 
\begin{equation}\label{rate_share}
\text{C4:}\quad \sum_{i\in\I(t)} c_i(t) \leq 1,
\end{equation}
and
\begin{equation}\label{rate_pos_c}
\text{C5:}\quad c_i(t) \geq 0,\,i\in\I(t).
\end{equation}

\textit{Remark 2:} 
The shared interests of the users can be exploited in the proposed the IRS-aided RS VR streaming system by properly choosing the values of $c_i(t),\,i\in\I(t),$ based on the video tile requests and CSI of the users.
Through the optimization of $c_i(t)$, the base station can determine which tiles should be included in the common message, and what are the corresponding proportions of the common rate that should be allocated to the transmission of these tiles.
When multiple users request the same tile, it may be beneficial to use a larger proportion of the common rate to transmit this tile since it can increase the individual utility of those users that requested this tile simultaneously.
The proposed algorithm for optimizing $c_i(t)$ along with the other DoF of the system will be presented in Sections III and IV.

\textit{Remark 3:} 
When $c_i=0$, this means that no data from tile $i$ is included in the common message.
As an example, for the FoVs and the corresponding video tile requests of the users shown in Fig. \ref{scenario_just}, one possible construction of the common message is given as follows: $c_{10}=c_{15}=c_{16}=\frac{1}{3}$, and $c_{9}=c_{11}=c_{14}=c_{17}=c_{20}=c_{21}=0$.
This means that the common message only includes data from tiles 10, 15, and 16, which are requested by multiple users, and the common rate is split equally between these three tiles.

After decoding the common message, user $n$ removes the signal corresponding to the common message from $y_n(t)$ using SIC, and decodes its private message by treating the private messages of other users as interference.
Thus, the SINR of the private message at user $n\in\N$ in time slot $t\in\T$ is given by \cite{mao2021rate}:
\begin{equation}\label{pri_sinr}
\begin{aligned}
&\gamma_n^{\text{p}}(t) \\
&= \frac{\big|\left(\h^H_{n,D}(t) + \h^H_{n,R}(t)\,\psm(t)\,\Gm(t)\right)\bfv_n(t)\big|^2}{\sum\limits_{m\in\N\setminus{\{n\}}} \hspace{-1mm} \big| \left(\h^H_{n,D}(t)+ \h^H_{n,R}(t)\,\psm(t)\,\Gm(t)\right)\bfv_m(t) \big|^2 + \sigma^2}.
\end{aligned}
\end{equation}
The achievable rate of the private message of user $n$ is given by $R^{\text{p}}_n(t)=\log_2(1+\gamma_n^{\text{p}}(t))$.
Let $p_{n,i}(t)$ denote the proportion of $R^{\text{p}}_n(t)$ that is used to transmit the data of tile $i\in\I_n(t)$ in time slot $t$.
We have 
\begin{equation}\label{pv_cons_1}
\text{C6:}\quad \sum_{i\in\I_n(t)} p_{n,i}(t) \leq 1,\,n\in\N,
\end{equation}
\begin{equation}\label{pv_cons_2}
\text{C7:}\quad p_{n,i}(t) \geq 0,\,i\in\I_n(t),\,n\in\N.
\end{equation}

\textit{Remark 4:} 
When $c_i(t)>0$ and $p_{n,i}(t)>0$, this means that portions of the data of tile $i\in\I_n(t)$ are included in both the common and private messages for user $n$ in time slot $t$.
However, when $c_i(t)=0$ and $p_{n,i}(t)>0$, the data of tile $i\in\I_n(t)$ is transmitted only via the private message to user $n$ in time slot $t$.

\subsection{Per-User Per-Tile QoS Requirement}
After decoding the common and private messages, user $n$ retrieves its requested video tiles by combining the decoded messages.
%In the common message, only those bits that belong to tile $i\in\I_n(t)$ are useful to user $n$.
%For the ease of presentation, we refer to these bits as the \textit{useful bits} of user $n$.
Let $J^{\ctxt}_{n,i}(t)$ denote the number of bits that user $n$ obtained from the common message for tile $i\in\I_n(t)$ in time slot $t$.
We have
\begin{equation}
J^{\ctxt}_{n,i}(t) = W \TD c_i(t) R^{\ctxt}(t),\,i\in\I_n(t),\,n\in\N,
\end{equation}
where $W$ and $\TD$ are the downlink transmission bandwidth and time duration, respectively.
Let $J^{\ptxt}_{n,i}(t)$ denote the number of bits that user $n$ obtained from its private message for tile $i\in\I_n(t)$ in time slot $t$.
We have 
\begin{equation}
J^{\ptxt}_{n,i}(t) = W \TD p_{n,i}(t) R_n^{\ptxt}(t),\,i\in\I_n(t),\,n\in\N.
\end{equation}
By combining the common and private messages, the total number of bits that user $n$ received for tile $i\in\I_n(t)$ in time slot $t$ is given by $
J_{n,i}(t) = J^{\text{c}}_{n,i}(t) + J^{\text{p}}_{n,i}(t),\,i\in\I_n(t),\,n\in\N$.
The total number of bits required by user $n$ to retrieve tile $i$ with bitrate $v_{n,i}(t)$ is given by $J^{\text{min}}_{n,i}(t) = T_v \,v_{n,i}(t),\,i\in\I_n(t),\,n\in\N$,
where $T_v$ denotes the time duration of a 360-degree video tile.
In order to ensure that all the data requested by user $n$ can be received within the downlink transmission window, we have the following per-user per-tile QoS constraint:
\begin{equation}\label{qoe_c}
\text{C8:}\quad J_{n,i}(t) \geq J^{\text{min}}_{n,i}(t),\,i\in\I_n(t),\,n\in\N.
\end{equation}

%The first part is obtained from decoding the common message and retrieving the data of its requested tiles within the common message.
%The second part is obtained from decoding its private message.
%Hence, the total achievable rate of user $n$ in time slot $t$ is given by $R_{n,tot}(t) = \sum_{i\in\I_n(t)} C_i(t)  + R_n(t)$.

\subsection{Problem Formulation}
%At the beginning of time slot $t\in\T$, we assume the sets $\I_n(t),\,n\in\N,$ are known to the base station after the tile requests of the users have been received.
%In addition, we assume the base station knows the bitrate selection of the previous time slot $t-1$, i.e., $\br_n(t-1),\,n\in\N$, for $t>1$.
In time slot $t$, we tackle the following utility maximization problem for an IRS-aided RS VR system:
\begin{equation}\label{qoe_pro}
\begin{aligned}
\underset{\substack{\B(t),\,\psm(t),\\\br_n(t),\,n\in\N,\\c_i(t),\,i\in\I(t),\\p_{n,i}(t),\,i\in\I_n(t),\,n\in\N}}{\text{maximize}} \quad &  u(t) \overset{\triangle}{=}\sum_{n\in\N} u_n(t)\\
\text{subject to } \quad & \text{constraints C1$-$C8},\\
& \text{C9:}\,\,\phi_l(t) \in [0,\,2\pi),\,l\in\{1,2,\ldots,L\},
\end{aligned}
\end{equation}
where constraint C1 ensures that the bitrate of each tile can only be chosen from set $\V$.
Constraint C2 is the maximum downlink transmission power constraint at the base station. 
Constraint C3 guarantees that the common message can be decoded by all users.
Constraints C4$-$C7 are the constraints for the common and private rates allocated to the requested video tiles.
%specify that the rates split among different tiles are always nonnegative and their summation does not exceed one.
%Constraint (\ref{td_cons}) ensures that the transmission delay does not exceed the maximum tolerable value.
%Constraints (\ref{pv_cons_1}) and (\ref{pv_cons_2}) specify that the portions of private rate allocated to different tiles are always nonnegative and their summation does not exceed the private rate.
Constraint C8 is the per-user per-tile QoS constraint.
Constraint C9 is the IRS phase shift constraint.
Problem (\ref{qoe_pro}) is a mixed-integer nonconvex optimization problem.
In the Appendix, we present an AO algorithm for solving problem (\ref{qoe_pro}) using FP, SDR, and convex optimization.
In the AO algorithm, we decompose problem (\ref{qoe_pro}) into three subproblems and solve them iteratively.
Although a suboptimal solution of problem (\ref{qoe_pro}) can be obtained with the AO algorithm, the AO algorithm can become computationally expensive and time-consuming for VR streaming applications, see the Appendix.
To tackle this issue, in the next section, we propose a learning-based Deep-GRAIL algorithm to efficiently solve problem (\ref{qoe_pro}).

\section{Deep-GRAIL: Deep Deterministic Policy Gradient with Imitation Learning Algorithm}
Designing a learning-based algorithm for solving problem (\ref{qoe_pro}) is challenging due to the RS encoding and decoding procedure, and the constraints in problem (\ref{qoe_pro}).
In the proposed Deep-GRAIL algorithm, we tackle these challenges using imitation learning \cite{nair2018demo,rashidinejad2021bridging}, and differentiable convex optimization \cite{agrawal2019}.

\subsection{Markov Decision Process (MDP) Formulation}\label{sec_ddpg}

We first model the sequential decision process for solving problem (\ref{qoe_pro}) in time slot $t\in\T$ as an MDP with $\tau^{\text{max}}$ decision epochs.
For notational simplicity, we drop time index $t$ in this section.
The state vector in the $\tau$-th decision epoch of the MDP is defined as
\begin{equation}\label{state_def}
\begin{aligned}
\sta(\tau) = \Big[&\h_{n,D},\,\text{vec}(\diag(\h^H_{n,R}) \Gm),\,n\in\N,\\
&\B(\tau-1),\,\text{vec}(\psm(\tau-1)),\,\,\boldsymbol{c}(\tau-1),\,R^\text{c}_n(\tau-1),\\
&R^\text{p}_n(\tau-1),\,\boldsymbol{p}_n(\tau-1),\,\br_n(\tau-1),\,\boldsymbol{o}_n,\,n\in\N\Big],
\end{aligned}
\end{equation}
where $\boldsymbol{p}_n(\tau) = (p_{n,i}(\tau),\,i\in\I)$
and $\boldsymbol{c}(\tau) = (c_{i}(\tau),\,i\in\I)$.
In addition, the binary vector $\boldsymbol{o}_n = (\mathds{1}\left(i\in\I_n\right),i\in\I)\in\{0,1\}^{I_{\text{max}}}$ in (\ref{state_def}) contains the information about the video tile request of user $n$.
Note that $\tau=0$ corresponds to the initialization of the MDP.

Furthermore, the action vector in the $\tau$-th decision epoch is defined as 
\begin{equation}
\act(\tau) = \left(\B(\tau),\,\text{vec}(\psm(\tau)),\boldsymbol{c}(\tau),\boldsymbol{p}_n(\tau),\br_n(\tau), n\in\N\right).
\end{equation}
The action vector $\act(\tau)$ contains all optimization variables of  problem (\ref{qoe_pro}).

For the reward function design, note that the objective function in problem (\ref{qoe_pro}), i.e., $u(\tau)$, can only take values from a finite set due to the discrete nature of bitrate selection.
If $u(\tau)$ is used directly as the reward function, then the reward function becomes sparse and may prevent the learning agent from effectively improving the policy \cite{andrychowicz2017hindsight}.
To tackle this issue, we first use the following inequality to establish the connection between $u(\tau)$ and the system sum-rate explicitly:
\begin{align}\label{obj_trans}
u(\tau) &= \sum_{n\in\N}\left(\sum_{i\in\I_n} v_{n,i}(\tau)-\kappa^{\text{intra}}\,\ell_n^{\text{intra}}(\tau)\right)\notag\\
&\overset{(a)}{\leq} \sum_{n\in\N} \frac{W \TD}{T_v} \Big(\sum_{i\in\I_n} p_{n,i}(\tau) R_n^{\ptxt}(\tau) + \sum_{i\in\I_n} c_i(\tau) R^{\ctxt}(\tau) \Big) \notag\\
& \hspace{5mm}- \sum_{n\in\N} \kappa^{\text{intra}}\,\ell_n^{\text{intra}}(\tau)\notag\\
&\overset{\triangle}= r(\sta(\tau),\act(\tau)),
\end{align}
where inequality $(a)$ follows from inequality (\ref{qoe_c}). 
That is
\begin{align}\label{qoe_c_trans}
&J_{n,i} \geq J^{\text{min}}_{n,i}(\tau)\notag\\
&\Longleftrightarrow W \TD \Big(\sum_{i\in\I_n} p_{n,i}(\tau) R_n^{\ptxt}(\tau) + \sum_{i\in\I_n} c_i(\tau) R^{\ctxt}(\tau) \Big)\notag\\
&\hspace{8mm}\geq T_v \sum_{i\in\I_n} v_{n,i}(\tau)\notag\\
&\Longrightarrow \sum_{n\in\N}\frac{W \TD}{T_v} \Big(\sum_{i\in\I_n} p_{n,i}(\tau) R_n^{\ptxt}(\tau) + \sum_{i\in\I_n} c_i(\tau)  R^{\ctxt}(\tau) \Big)\notag\\
&\hspace{8mm}\geq \sum_{n\in\N}\sum_{i\in\I_n} v_{n,i}(\tau).
\end{align}
We use $r(\sta(\tau),\act(\tau))$ as the reward function in the MDP to provide an informative feedback to the learning agent.
In the reward function $r(\sta(\tau),\act(\tau))$, we replace the discrete bitrate selections with the continuous achievable bitrates to overcome the sparsity of $u(\tau)$.
We set the maximum achievable bitrate of a video tile in $r(\sta(\tau),\act(\tau))$ to be equal to the maximum bitrate selection $v_M$.
%such that the learning agent will not receive additional rewards by allocating resources to those users that can obtain the requested tiles with the maximum bitrate selection $v_M$.
Since the difference between $r(\sta(\tau),\act(\tau))$ and $u(\tau)$ cannot exceed $\sum_{n\in\N} |\I_n| \max_{i=1,\ldots,M-1} |v_{i+1} - v_i|$, using $r(\sta(\tau),\act(\tau))$ as the reward function also leads to a policy that can achieve a high utility $u(\tau)$.

\subsection{Actor-Critic Method with $q$-Step Return}

Based on the MDP formulation, we use an actor network to learn a policy for solving problem (\ref{qoe_pro}).
We denote the learnable parameters (i.e., the weights and biases) of the actor network as $\TA_{\text{act}}$.
The policy learned by the actor network, which is denoted by $\pi_{\TA_{\text{act}}}(\sta(\tau))$, defines a mapping from a state to an action. 
That is, $\act(\tau) = \pi_{\TA_{\text{act}}}(\sta(\tau))$.
The critic network learns a state-action value function $Q_{\TA_{\text{crt}}}$, which is parameterized by $\TA_{\text{crt}}$.
The state-action value function $Q_{\TA_{\text{crt}}}(\sta(\tau),\act(\tau))$ estimates the discounted total reward of selecting action $\act(\tau)$ in state $\sta(\tau)$.
That is,
\begin{align}\label{Qfunc_def}
&Q_{\TA_{\text{crt}}}(\sta(\tau),\act(\tau))\nonumber \\
&= \mathbb{E}_{\sta \sim p_{\pi_{\TA_{\text{act}}}}, \act \sim \pi_{\TA_{\text{act}}}}\left[\sum_{\tau^\prime=\tau}^{\tau^{\text{max}}} \gamma^{\tau^\prime-\tau} r(\sta(\tau^\prime),\act(\tau^\prime))\right],
\end{align}
where $p_{\pi_{\TA_{\text{act}}}}$ denotes the distribution of the state transition as the result of taking actions based on policy $\pi_{\TA_{\text{act}}}$, and $\gamma\in[0,1]$ is the discount factor.

The goal of the actor-critic method is to learn a policy which maximizes the discounted total reward \cite{timothy2016ddpg}. 
We have
\begin{equation}\label{ac_opt}
\begin{aligned}
&\underset{\substack{\TA_{\text{act}}}}{\text{maximize}}\; \mathcal{L}(\TA_{\text{act}})\\
&\triangleq \mathbb{E}_{\sta \sim p_{\pi_{\TA_{\text{act}}}}, \act \sim \pi_{\TA_{\text{act}}}}\left[\sum_{\tau^\prime=1}^{\tau^{\text{max}}} \gamma^{\tau^\prime-1} r(\sta(\tau^\prime),\act(\tau^\prime))\right].
\end{aligned}
\end{equation}
The deterministic policy gradient \cite{timothy2016ddpg} for solving problem (\ref{ac_opt}) is given by:
\begin{equation}\label{dpg_gradient}
\nabla \mathcal{L}_{\text{DPG}} = \mathbb{E}_{\sta \sim p_{\pi_{\TA_{\text{act}}}}}\left[\nabla Q_{\TA_{\text{crt}}}(\sta,\act)\,|\,_{\act=\pi_{\TA_{\text{act}}}(\sta)} \nabla \pi_{\TA_{\text{act}}}(\sta) \right].
\end{equation}

The policy update based on the deterministic policy gradient may suffer from  Q-value overestimation \cite{TD32018scott}.
To tackle the overestimation issue in the proposed Deep-GRAIL algorithm, we use the following two techniques. 
First, while the vanilla DDPG algorithm \cite{timothy2016ddpg} only uses one critic network, we use $V$ critic networks to obtain $V$ independent approximations of the Q-value.
The target of Q-value approximation is determined by the minimum of these $V$ approximations to alleviate overestimation. 
Second, we determine the target of Q-value approximation using the $q$-step return \cite[Section~7.1]{sutton1998introduction}.
Compared with the single-step return, $q$-step return determines the discounted future reward over $q$ consecutive decision epochs.
It provides the learning agent with more information regarding future planning when compared with the single-step return (i.e., when $q=1$).
Hence, the target Q-value in the proposed Deep-GRAIL algorithm is given by:
\begin{equation}\label{tar_Q_value}
\begin{aligned}
&\widehat{Q}_{\TA_{\text{crt}}}(\sta(\tau), \pi_{\TA_{\text{act}}}(\sta(\tau)) \\
&=  \sum_{\tau^\prime=\tau}^{\tau+q-1} \gamma ^{\tau^\prime - \tau} r(\sta(\tau^\prime),\pi_{\TA_{\text{act}}}(\sta(\tau^\prime))) \\
&\hspace{5 mm}+ \gamma^q\, \min_{m=1,2,\ldots,V} Q_{\TA^{(m)}_{\text{crt}}}(\sta(\tau+q),\pi_{\TA_{\text{act}}}(\sta(\tau+q))),
\end{aligned}
\end{equation}
where $\TA^{(m)}_{\text{crt}}$ denotes the learnable parameters of the $m$-th critic network.
Then, $\TA^{(m)}_{\text{crt}},m=1,\ldots,V$, is updated by minimizing the following temporal difference (TD) error of Q-value approximation \cite[Ch.~6]{sutton1998introduction}:
\begin{align}
&\underset{\substack{\TA^{(m)}_{\text{crt}}}}{\text{minimize}}\;\mathcal{L}(\TA^{(m)}_{\text{crt}}) \nonumber\\
&\triangleq \mathbb{E}_{\sta \sim p_{\pi_{\TA_{\text{act}}}}, \act \sim \pi_{\TA_{\text{act}}}}\Big[ \big( Q_{\TA^{(m)}_{\text{crt}}}(\sta,\pi_{\TA_{\text{act}}}(\sta)) \nonumber\\
&\hspace{50mm}-  \widehat{Q}_{\TA_{\text{crt}}}(\sta, \pi_{\TA_{\text{act}}}(\sta))\big)^2\Big]\nonumber.
\end{align}
We update $\TA^{(m)}_{\text{crt}},m=1,\ldots,V,$ using gradient descent with the following gradient:
\begin{equation}\label{crt_gradient}
\begin{aligned}
&\nabla \mathcal{L}(\TA^{(m)}_{\text{crt}}) \\
&= \mathbb{E}_{\sta \sim p_{\pi_{\TA_{\text{act}}}}, \act \sim \pi_{\TA_{\text{act}}}}\bigg[ 2\nabla Q_{\TA^{(m)}_{\text{crt}}}(\sta,\pi_{\TA_{\text{act}}}(\sta)) \\
&\hspace{22mm}\big( Q_{\TA^{(m)}_{\text{crt}}}(\sta,\pi_{\TA_{\text{act}}}(\sta)) -  \widehat{Q}_{\TA_{\text{crt}}}(\sta, \pi_{\TA_{\text{act}}}(\sta))\big)\bigg].
\end{aligned}
\end{equation}

The learning agent maintains an experience replay that stores the system transition history due to past decisions as a \textit{system transition tuple} $(\sta(\tau),\act(\tau), r(\sta(\tau),\act(\tau)), \sta(\tau+1))$.
To determine the gradients in (\ref{dpg_gradient}) and (\ref{crt_gradient}), we first sample the system transition tuples of $q$ consecutive decision epochs from the experience replay, and then determine the gradients for the sampled system transition.
Note that $q$ consecutive decision epochs are sampled in order to determine the $q$-step return in (\ref{tar_Q_value}).
We find the gradients in a minibatch-based manner and average the gradients over $M_D$ different samples.

Note that while the aforementioned actor-critic method is designed for maximizing the expected discounted total reward, we tackle the constraints in problem (\ref{qoe_pro}) during the learning process by using DCO layers in the DNN structure design of the proposed RavNet.
The corresponding details will be presented in Section \ref{sec_dnn}.

\subsection{Policy Improvement using Imitation Learning and Demonstration Replay}
In the vanilla DDPG algorithm \cite{timothy2016ddpg} and its variant \cite{TD32018scott}, the learning agent improves the learned policy only based on exploration and experience replay.
However, the exploration of the learning agent can be inefficient when the dimensionality of the state and action space is large.
To tackle this issue, we notice that problem (\ref{qoe_pro}) has a hidden convexity due to the fractional form of the SINR expressions in (\ref{common_sinr}) and (\ref{pri_sinr}).
Hence, a suboptimal solution of problem (\ref{qoe_pro}) can be obtained by using an AO algorithm exploiting convex optimization, FP, and SDR.
The details of the AO algorithm are provided in the Appendix.

In the proposed Deep-GRAIL algorithm, we use imitation learning  \cite{rashidinejad2021bridging,nair2018demo} to allow the agent to learn from the AO algorithm and exploit the hidden convexity of problem (\ref{qoe_pro}).
We introduce the \textit{imitation loss} to characterize the difference between the action chosen by the learned policy and the suboptimal solution obtained by the AO algorithm.
By using imitation learning, the proposed Deep-GRAIL algorithm provides the actor-critic method with knowledge about the hidden convexity of the formulated problem, leading to an efficient policy learning.

\begin{figure}[!t]
\centering
\includegraphics[width=3.2in]{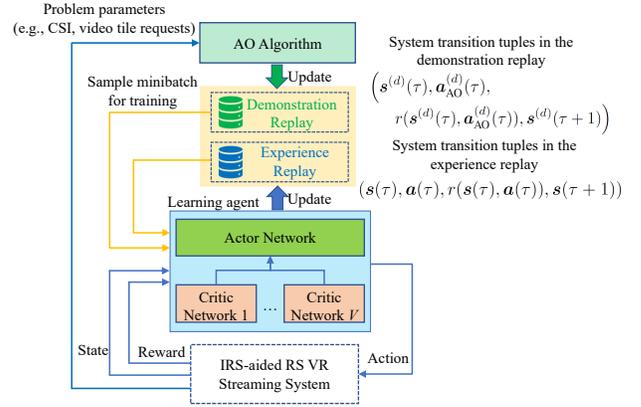}
%\vspace{-5mm}
\caption{Overall framework of the proposed Deep-GRAIL algorithm.
The learning agent comprises an actor network and $V$ critic networks.
An experience replay is employed to maintain the exploration history of the learning agent.
A demonstration replay is maintained by the learning agent to store the system transition tuples obtained from the AO algorithm. 
}
\label{imitationRL}
%\vspace{-2mm}
\end{figure}

In particular, the framework of the proposed Deep-GRAIL algorithm is illustrated in Fig. \ref{imitationRL}.
Apart from the actor-critic method and the experience replay, we introduce a \textit{demonstration replay} to store the solutions of problem (\ref{qoe_pro}) obtained by the AO algorithm over $D$ time slots.
Note that one time slot corresponds to one episode comprising $\tau^{\text{max}}$ decision epochs in the MDP.
We define set $\D=\{1,2,\ldots,D\}$.
In the $d$-th time slot, where $d\in\D$, the AO algorithm is invoked to solve problem (\ref{qoe_pro}).
Since AO leads to an iterative algorithm, using it to solve problem (\ref{qoe_pro}) also results in a sequential system transition.
In particular, in the $\tau$-th decision epoch of the $d$-th time slot, we first initialize the AO algorithm with the variables in $\sta^{(d)}(\tau)$, which is the state vector in the $\tau$-th decision epoch of the $d$-th time slot.
Here, we use the superscript $(d)$ to denote the values of variables in the $d$-th time slot.
We then execute the AO algorithm for one iteration and obtain the solution.
We denote this solution as $\act^{(d)}_{\text{AO}}(\tau)$.
With $\act^{(d)}_{\text{AO}}(\tau)$, we determine the reward and the next state of the MDP as $r(\sta^{(d)}(\tau),\act^{(d)}_{\text{AO}}(\tau))$ and $\sta^{(d)}(\tau+1)$, respectively.
Then, the system transition obtained from the execution of the AO algorithm in the $\tau$-th decision epoch of the $d$-th time slot is denoted as the system transition tuple
$(\sta^{(d)}(\tau), \act^{(d)}_{\text{AO}}(\tau), r(\sta^{(d)}(\tau), \act^{(d)}_{\text{AO}}(\tau)),\sta^{(d)}(\tau+1))$.
We index this system transition tuple with the tuple $(d,\tau)$.
The system transition tuples obtained from the execution of the AO algorithm are stored in the demonstration replay for imitation learning.

In each training iteration, we sample a minibatch of $M_D$ different transition tuples from the demonstration replay.
We denote the set that collects the indices of the system transition tuples within the minibatch as $\M_D$.
Then, for each state $\sta^{(d)}(\tau),(d,\tau)\in\M_D,$ in the minibatch, we determine the imitation loss of the actor network based on the mean squared error between the action chosen by the actor network, i.e., $\pi_{\TA_{\text{act}}}(\sta^{(d)}(\tau))$, and the solution obtained by the AO algorithm, i.e., $\act^{(d)}_{\text{AO}}(\tau)$.
In particular, the imitation loss $\widehat{\mathcal{L}}_{\text{IMI}}$ is given by:
\begin{equation}\label{imitation_loss_ori}
\widehat{\mathcal{L}}_{\text{IMI}} = \frac{1}{M_D}\sum_{(d,\tau)\,\in\M_D} || \pi_{\TA_{\text{act}}}(\sta^{(d)}(\tau)) - \act^{(d)}_{\text{AO}}(\tau) ||^2.
\end{equation}
We scale the elements in $\pi_{\TA_{\text{act}}}(\sta^{(d)}(\tau))$ and $\act^{(d)}_{\text{AO}}(\tau)$ that correspond to beamforming and IRS phase shift variables to be between $-1$ and $1$ to mitigate the potential impact of the different ranges of the variables.

Note that the solution obtained by the AO algorithm is in general suboptimal due to the nonconvexity of the formulated problem.
Hence, using the imitation loss in (\ref{imitation_loss_ori}) may prevent the learning agent from finding better solutions than those obtained by the AO algorithm.
To tackle this issue, for each sample with index $(d,\tau)\in\M_D$, we determine the discounted total reward that can be achieved by following the actor's policy in the remaining decision epochs (i.e., the Monte Carlo estimation of the value function \cite[Section~7.1]{sutton1998introduction}) by
\begin{equation}
\widehat{Q}^{(d)}_{\TA_{\text{act}}}(\tau) = \sum_{\tau^\prime = \tau}^{\tau^{\text{max}}} \gamma^{\tau^\prime-\tau} r(\sta^{(d)}(\tau^\prime), \pi_{\TA_{\text{act}}}(\sta^{(d)}(\tau^\prime))).
\end{equation} 
The discounted total reward obtained by using the AO algorithm in the remaining decision epochs is given by
\begin{equation}
\widehat{Q}^{(d)}_{\text{AO}}(\tau) = \sum_{\tau^\prime = \tau}^{\tau^{\text{max}}} \gamma^{\tau^\prime-\tau} r(\sta^{(d)}(\tau^\prime),\act^{(d)}_{\text{AO}}(\tau^\prime)).
\end{equation}
To overcome the potential suboptimality of the AO algorithm, we only calculate the imitation loss for those states and actions for which the AO algorithm achieves a higher discounted total reward than the actor's policy.
This results in the following modified imitation loss:
\begin{equation}\label{imitation_loss}
\begin{aligned}
\mathcal{L}_{\text{IMI}} = \frac{1}{\widehat{M}_D} \sum_{(d,\tau)\in\M_D} \Bigg(&|| \pi_{\TA_{\text{act}}}(\sta^{(d)}(\tau)) - \act^{(d)}_{\text{AO}}(\tau) ||^2 \\
&\mathds{1}\left(\widehat{Q}^{(d)}_{\text{AO}}(\tau) > \widehat{Q}^{(d)}_{\TA_{\text{act}}}(\tau)\right)\Bigg),
\end{aligned}
\end{equation}
where $\widehat{M}_D = \sum_{(d,\tau)\in\M_D} \mathds{1}\left(\widehat{Q}^{(d)}_{\text{AO}}(\tau) > \widehat{Q}^{(d)}_{\TA_{\text{act}}}(\tau)\right)$.

By combining with the deterministic policy gradient in (\ref{dpg_gradient}), the overall gradient for updating the actor network in the proposed Deep-GRAIL algorithm is given by:
\begin{equation}\label{actor_gradient}
\nabla \mathcal{L}_{\text{act}} = \omega_1  \nabla \mathcal{L}_{\text{DPG}} + \omega_2 \nabla \mathcal{L}_{\text{IMI}},
\end{equation}
where $\omega_1$ and $\omega_2$ are positive parameters representing the weights of the deterministic policy gradient and the gradient of imitation loss, respectively.

\subsection{Training Algorithm}
\begin{algorithm}[!t]
%\algsetup{linenosize=\small}
%\small
\caption{Deep-GRAIL: Training Algorithm}\label{training_algo}
\begin{algorithmic}[1]
\STATE Set episode counter $t \leftarrow 0$.
\STATE Initialize $\TA_{\text{act}}$, $\TA^{(m)}_{\text{crt}},m=1,\ldots,V$.
\STATE Execute the AO algorithm for $D$ episodes and store the system transition tuples in the demonstration replay.
\STATE Perform random exploration for $T_{\text{warm-up}}$ episodes and store the system transition tuples in the experience replay.
\WHILE{$t \leq T_{\text{max}}$}
\STATE{Observe the CSI and $\I_n(t)$ of the users.}
\STATE Initialize $c_i(0) = \mathds{1}(i\in\I(t)) \frac{1}{|\I(t)|}$, $p_{n,i}(0)=\mathds{1}(i\in\I_n(t)) \frac{1}{I_n(t)}$, $i\in\I, n\in\N$.
\STATE Initialize $\psm(0)$ and $\B(0)$ based on random initialization.
\STATE Initialize $\tau \leftarrow 1$.
\WHILE{$\tau \leq \tau^{\text{max}}$}
\STATE Determine the action $\act(\tau) \leftarrow \pi_{\TA_{\text{act}}}(\sta(\tau))\,+\,\boldsymbol{\varrho}_{\text{epl}}$.
\STATE Observe the reward $r(\sta(\tau),\act(\tau))$.
\STATE Obtain the next state $\sta(\tau+1)$ and store the tuple $(\sta(\tau),\act(\tau),r(\tau),\sta(\tau+1))$ in the experience replay.
\STATE Sample $M_D$ transition tuples from the demonstration replay and experience replay, respectively.
\STATE Determine the gradients for updating the actor and critic networks based on (\ref{actor_gradient}) and (\ref{crt_gradient}), respectively.
\STATE Update $\TA_{\text{act}}$, $\TA^{(m)}_{\text{crt}},m=1,\ldots,V,$ using the Adam optimizer and (\ref{soft_update}). 
\STATE $\tau \leftarrow \tau+1$.
\ENDWHILE 
\STATE $t \leftarrow t+1$.
\ENDWHILE 
\end{algorithmic}
\end{algorithm}
The proposed training algorithm is illustrated in \textbf{Algorithm \ref{training_algo}}.
%One time slot corresponds to one episode with $\tau^{\text{max}}$ decision epochs.
We first obtain the demonstration replay by executing the AO algorithm over $D$ episodes.
This results in $D \tau^{\text{max}}$ system transition tuples being stored in the demonstration replay.
Meanwhile, the learning agent explores the state and action space by taking actions based on the learned policy, and stores the resulting system transition tuples in the experience replay.

We train the learning agent for $T_{\text{max}}$ episodes.
In each training iteration, we first sample a minibatch from the demonstration replay and determine the imitation loss based on (\ref{imitation_loss}).
Then, we sample another minibatch from the experience replay and determine the gradient for updating $\TA^{(m)}_{\text{crt}},m=1,\ldots,V$, based on (\ref{crt_gradient}).
Moreover, we determine the gradient for updating the actor network based on  (\ref{actor_gradient}).
We use the Adam optimizer \cite{kingma2015adam} with a learning rate of $\alpha$ to update the learnable parameters of the actor and critic networks based on the aforementioned gradients.
In addition, the following techniques are employed in our training algorithm to improve the efficiency of policy learning:
\begin{itemize}
\item \textit{Exploration noise}: We add an exploration noise $\boldsymbol{\varrho}_{\text{epl}}$ to the action determined by the actor network during the training phase to facilitate the exploration of the learning agent.
The elements in $\boldsymbol{\varrho}_{\text{epl}}$ are generated from the Gaussian distribution with zero mean and variance $\sigma^2_{\text{epl}}$\footnote{\textcolor{black}{Note that we scale the values of the elements in $\act(\tau)$ that correspond to the beamforming and IRS phase shift variables to be between $-1$ and $1$ in our implementation.
Hence, we can generate the exploration noise for all elements in $\act(\tau)$ using the same Gaussian distribution since they all have the same range of magnitudes.}}.
In addition, the learning agent randomly explores the state and action spaces for $T_{\text{warm-up}}$ episodes before updating the learnable parameters.

%That is, the action chosen during the training phase is given by
%\begin{equation}\label{action_explore}
%\act(\tau) \leftarrow \pi_{\TA_{\text{act}}}(\sta(\tau))\,+\,\boldsymbol{\varrho}_{\text{epl}}.
%\end{equation}

\item \textit{Delayed actor network update}: Delaying the update of the actor network can alleviate the impact of overestimation of the critic network on policy learning \cite{TD32018scott}.
In the proposed Deep-GRAIL algorithm, we update the actor network every $\delta$ training iterations ($\delta>1$), while the critic networks are updated in each iteration.

\item \textit{Soft learnable parameter update}: The soft update technique can stabilize the training process and prevent divergence.
Let $\TA^{(m)'}_{\text{crt}},m=1,\ldots,V,$ and $\TA'_{\text{act}}$ denote the new parameters of the critic and actor networks that we obtain based on the gradients in (\ref{crt_gradient}) and (\ref{actor_gradient}), respectively. 
We update the learnable parameters with the following soft update rule:
\begin{equation}\label{soft_update}
\begin{aligned}
\TA^{(m)}_{\text{crt}} &\leftarrow \kappa \TA^{(m)'}_{\text{crt}} + (1-\kappa)\TA^{(m)}_{\text{crt}},m=1,\ldots,V,\\
\TA_{\text{act}} &\leftarrow \kappa \TA'_{\text{act}} + (1-\kappa)\TA_{\text{act}},
\end{aligned}
\end{equation}
\hspace{-2mm}where $\kappa$ is a constant and is between zero and one. 
\end{itemize}

%With $\boldsymbol{c}(\tau), \boldsymbol{p}_n(\tau)$, and $\br(\tau)$, we obtain the next state, i.e., $\sta(\tau+1)$, and repeat the aforementioned process iterative until the maximum decision step is reached. 

\subsection{Online Execution Algorithm}
The algorithm for online execution is illustrated in \textbf{Algorithm \ref{online_algo}}.
Given the CSI and video tile requests of the users, the proposed Deep-GRAIL algorithm is executed for $\tau^{\text{max}}$ iterations to obtain the solutions.
In the $\tau$-th iteration, we feed state $\sta(\tau)$ into the actor network and determine the action $\act(\tau)$.
Then, the next state $\sta(\tau+1)$ is observed.
We repeat this process iteratively until the maximum decision epoch $\tau^{\text{max}}$ is reached.
Compared with the training algorithm, the online execution algorithm has a lower computational complexity since the update of the learnable parameters and the demonstration replay (i.e., the execution of the AO algorithm) are not required during online execution.

\section{RavNet: Proposed Deep Neural Network for IRS-aided RS VR Streaming Systems}
In this section, we propose RavNet, a DNN that we design for policy learning in the considered RS VR streaming
system.
With the help of the DCO layer, we are able to integrate convex optimization as one of the DNN layers in RavNet.
When combined with the proposed Deep-GRAIL algorithm, RavNet is capable of learning the policy efficiently, and meanwhile satisfying the constraints in problem (\ref{qoe_pro}).
\subsection{Input Pre-processing}
We first construct two three-dimensional (3-D) matrices, namely $\Sta^{(1)}(\tau)$ and $\Sta^{(2)}(\tau)$, from state $\sta(\tau)$.
$\Sta^{(1)}(\tau)$ collects the information about the channel, beamforming vectors, and IRS phase shifts. 
$\Sta^{(2)}(\tau)$ collects the information about the video tile requests, RS parameters, and bitrate selections.
\subsubsection{Construction of $\Sta^{(1)}(\tau)$}
$\Sta^{(1)}(\tau)$ is a 3-D matrix of size $(L + 1) N_t \times N \times 7$.
For ease of presentation, we refer to the first, second, and third dimensions of the 3-D matrix as row, column, and depth, respectively.
$\Sta^{(1)}[x,y,z]$ returns the element in the $x$-th row, $y$-th column, and $z$-th depth of matrix $\Sta^{(1)}(\tau)$.
$\Sta^{(1)}[:,y,z],z=1,\ldots,7,$ returns the $y$-th column vector in the $z$-th depth of $\Sta(\tau)$.
$\Sta^{(1)}[:,:,z],z=1,\ldots,7,$ returns the 2-D matrix of size $(L + 1) N_t \times N$ in the $z$-th depth of $\Sta^{(1)}$.
The elements in the first and second depths of $\Sta^{(1)}(\tau)$ are constructed from the real and imaginary parts of the CSI of the users, respectively.
We have 
\begin{equation}
\begin{aligned}\nonumber
\Sta^{(1)}[:,n,1](\tau) = \Big(&\Re\left\{\h_{n,D}\right\}, \\
&\Re\left\{\text{vec}(\diag(\h^H_{n,R}) \Gm)\right\}\Big),\,n\in\N.
\end{aligned}
\end{equation}
and
\begin{equation}
\begin{aligned}\nonumber
\Sta^{(1)}[:,n,2](\tau) = \Big(&\Im\left\{\h_{n,D}\right\}, \\
&\Im\left\{\text{vec}(\diag(\h^H_{n,R}) \Gm)\right\}\Big),\,n\in\N.
\end{aligned}
\end{equation}
\begin{figure*}[!t]
\centering
\includegraphics[width=5.6in]{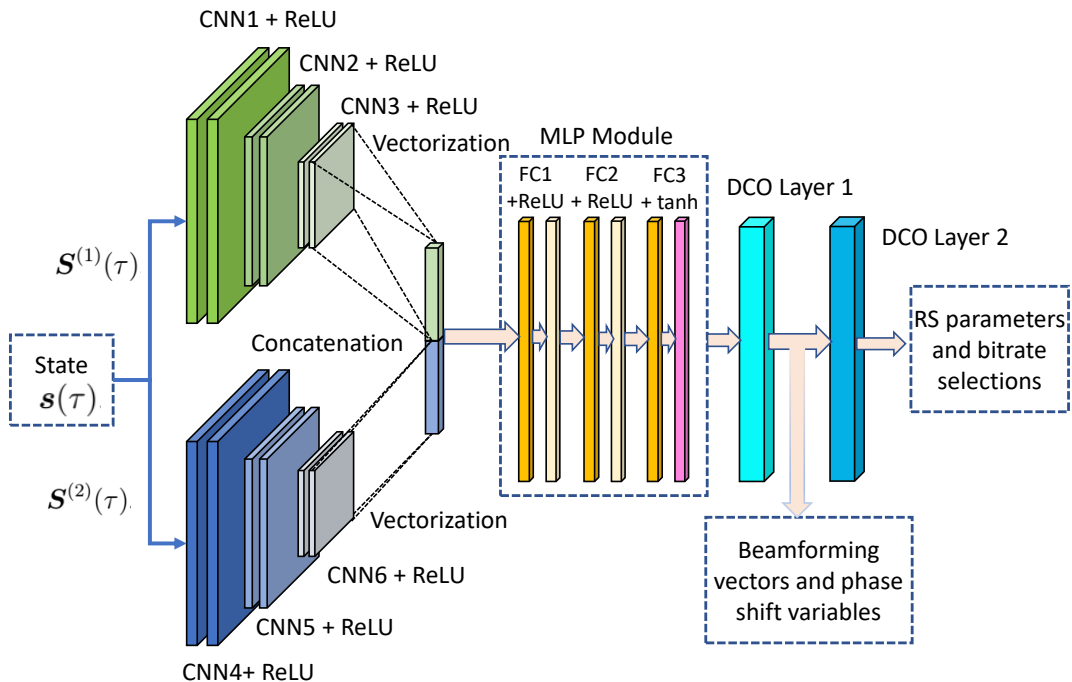}
%\vspace{-5mm}
\caption{The network architecture of the actor network in RavNet. The actor network takes $\Sta^{(1)}(\tau)$ and $\Sta^{(2)}(\tau)$ as input, and determines the action $\act(\tau)$.
}
\label{DNNStructure}
\end{figure*}

\begin{algorithm}[!t]
\caption{Deep-GRAIL: Online Execution Algorithm for Time Slot $t$}\label{online_algo}
\begin{algorithmic}[1]
%\small
\STATE{Obtain the CSI and video tile requests of the users.}
%\STATE Conduct $\tau_{\text{warm-up}}$ episodes of exploration and store the transition history.
\STATE Initialize $c_i(0) = \mathds{1}(i\in\I(t)) \frac{1}{|\I(t)|}$, $p_{n,i}(0)=\mathds{1}(i\in\I_n(t)) \frac{1}{I_n(t)}$, $i\in\I, n\in\N$.
\STATE Initialize $\psm(0)$ and $\B(0)$ based on random initialization.
\STATE Initialize $\tau \leftarrow 1$.
\WHILE{$\tau \leq \tau^{\text{max}}$}
\STATE Determine the action $\act(\tau) \leftarrow \pi_{\TA_{\text{act}}}(\sta(\tau))$.
\STATE Obtain the next state $\sta(\tau+1)$.
\STATE $\tau \leftarrow \tau+1$.
\ENDWHILE 
\STATE Retrieve $\B(t)$, $\psm(t)$, $\boldsymbol{c}(t)$, $\boldsymbol{p}(t),$ and $\br(t)$ from $\act(\tau^{\text{max}})$.
\end{algorithmic}
\end{algorithm}
The remainder of $\Sta^{(1)}(\tau)$ is constructed from the beamforming vectors and IRS phase shifts.
We align the beamforming vectors and IRS phase shifts with the CSI of the corresponding subchannels in the depth dimension of matrix $\Sta^{(1)}(\tau)$.
By doing this, we allow the DNN to learn from the \textit{positional information} (e.g., each beamforming or IRS phase shift  variable is linked to the corresponding subchannel) in matrix $\Sta^{(1)}(\tau)$.

In particular, the elements in the 3rd and $4$th depth of $\Sta^{(1)}(\tau)$ are obtained from the beamforming vector of the common message chosen in the previous decision epoch, i.e., $\bfv_0(\tau-1)$.
For $n\in\N$, we have
\begin{equation}\nonumber
\begin{aligned}
\Sta^{(1)}[:,n,3](\tau) = \Big(\underbrace{\Re\left\{\bfv_0(\tau-1)\right\},\ldots,\Re\left\{\bfv_0(\tau-1)\right\}}_{L+1}\Big),
\end{aligned}
\end{equation}
and 
\begin{equation}\nonumber
\Sta^{(1)}[:,n,4](\tau) = \Big(\underbrace{\Im\left\{\bfv_0(\tau-1)\right\},\ldots,\Im\left\{\bfv_0(\tau-1)\right\}}_{L+1}\Big).
\end{equation}

The elements in the $5$th and $6$th depth of $\Sta^{(1)}(\tau)$ are obtained from the beamforming vector of the private messages chosen in the previous decision epoch, i.e., $\bfv_n(\tau-1)$.
For $n\in\N$, we have
\begin{equation}\nonumber
\Sta^{(1)}[:,n,5](\tau) = \Big(\underbrace{\Re\left\{\bfv_n(\tau-1)\right\},\ldots,\Re\left\{\bfv_n(\tau-1)\right\}}_{L+1}\Big),
\end{equation}
and 
\begin{equation}\nonumber
\Sta^{(1)}[:,n,6](\tau) = \Big(\underbrace{\Im\left\{\bfv_n(\tau-1)\right\},\ldots,\Im\left\{\bfv_n(\tau-1)\right\}}_{L+1}\Big).
\end{equation}

The elements in the last depth of $\Sta^{(1)}(\tau)$ are determined by the IRS phase shifts chosen in the previous decision epoch.
For $n\in\N$, we have
\begin{equation}\nonumber
\begin{aligned}
\Sta^{(1)}[:,n,7](\tau) = \Big(&\underbrace{0,\ldots,0}_{N_t}, \underbrace{\psi_1(\tau-1),\ldots,\psi_1(\tau-1)}_{N_t},\\
&\ldots,\underbrace{\psi_{L}(\tau-1),\ldots,\psi_{L}(\tau-1)}_{N_t}\Big).
\end{aligned}
\end{equation}

\subsubsection{Construction of $\Sta^{(2)}(\tau)$}
$\Sta^{(2)}(\tau)$ is a 3-D matrix of size $N_x \times N_y \times 2N$.
We construct $\Sta^{(2)}(\tau)$ in such a way that the 2-D matrices $\Sta^{(2)}[:,:,n]$ and $\Sta^{(2)}[:,:,n+N]$ show the achievable bitrates of the video tiles requested by user $n\in\N$ obtained from the common and private messages, respectively, based on the control variables determined in the previous decision epoch.
We have
\begin{equation}\nonumber
\begin{aligned}
&\Sta^{(2)}[x,y,z](\tau) \\
&= \begin{cases}
\mathds{1}\left\{i \in\I_z\right\} c_{i}(\tau-1) R^\text{c}(\tau-1)),\;\;\;\;z=1,\ldots,N,\\
\mathds{1}\left\{i \in\I_{z-N}\right\} p_{z-N, i}(\tau-1) R_{z-N}^\text{p}(\tau-1),\\
\hspace{45mm}z=N+1,\ldots,2N,
\end{cases}
\end{aligned}
\end{equation}
where $i = y+(x-1)N_x$ for $x=1,\ldots N_x$, and $y=1,\ldots,N_y$. 
\textcolor{black}{
In fact, the value of $i$ corresponds to the index of the video tile located in the $x$-th row and the $y$-th column of the 360-degree video frame.
}
\subsection{Actor Network Structure}
\label{sec_dnn}
The actor network takes both $\Sta^{(1)}(\tau)$ and $\Sta^{(2)}(\tau)$ as inputs to determine the control variables.
The proposed actor network structure tackles the constraints of problem (\ref{qoe_pro}) during the policy learning process. 
As shown in Fig. \ref{DNNStructure}, we use the following DNN modules in the actor network:
\subsubsection{Convolutional Neural Network (CNN) Module}
First, $\Sta^{(1)}(\tau)$ is fed into three CNN layers with kernel sizes $k_1\times k_1$, $k_2 \times k_2$, $k_3 \times k_3$, and channel numbers $ch_1$, $ch_2$, and $ch_3$, respectively.
Each CNN layer is followed by a rectified linear unit (ReLU) activation layer.
The output of the last CNN layer is vectorized into a vector, which is denoted by $\sta^{(1)}(\tau)$.

We employ another network module comprising three CNN layers to process $\Sta^{(2)}(\tau)$.
The kernel sizes of these three CNN layers are given by $k_4\times k_4$, $k_5\times k_5$, and $k_6\times k_6$, while their channel numbers are $ch_4$, $ch_5$, and $ch_6$, respectively.
We also apply an ReLU activation layer after each of the CNN layers.
The output of the last CNN layer is reshaped into a vector, which we denote as $\sta^{(2)}(\tau)$.

The CNN module is an important component in the proposed actor network because (a) it can learn from the \textit{positional information} in $\Sta^{(1)}(\tau)$, and (b) it can capture the spatial correlation between the video tile requests of the users in $\Sta^{(2)}(\tau)$.

\subsubsection{Multilayer Perceptron (MLP) Module}
We concatenate $\sta^{(1)}(\tau)$ and $\sta^{(2)}(\tau)$ together to obtain a new vector $\sta^{(3)}(\tau)$.
That is, $\sta^{(3)}(\tau) =(\sta^{(1)}(\tau), \sta^{(2)}(\tau))$.
We feed $\sta^{(3)}(\tau)$ into an MLP module with three fully-connected (FC) layers, two ReLU activation layers, and one tanh activation layer to obtain the beamforming variables $\B^\prime(\tau)$ and IRS phase shifts $\psmv^\prime(\tau)$.
We denote the hidden dimensionality of the MLP module in the actor network as $d_{\text{act}}$.
% \textcolor{blue}{
% We denote the sizes of the three FC layers in the actor network as $d_{\text{act1}}\times  d_{\text{act2}}$, $d_{\text{act2}} \times d_{\text{act3}}$, and $d_{\text{act3}}\times d_{\text{act\_out}}$, respectively.
% }
We define $\act^\prime(\tau) = (\B^\prime(\tau),\psmv^\prime(\tau))$.
 
\subsubsection{DCO Layers}
In order to satisfy the constraints in problem (\ref{qoe_pro}), we determine the projection of $\act^\prime(\tau)$ onto the feasible set of problem (\ref{qoe_pro}) by solving the following optimization problem with the RS parameters and bitrate selection given by $\boldsymbol{c}(\tau-1)$, $\boldsymbol{p}_n(\tau-1)$, and $\br_n(\tau-1)$, respectively:
\begin{equation}\label{safe_rl_pro}
\begin{aligned}
\underset{\substack{\act(\tau)}}{\text{minimize}} \quad &  ||\act(\tau)-\act^\prime(\tau)||^2\\
\text{subject to } \quad & \text{constraints C2, C8, C9}.
\end{aligned}
\end{equation}
Note that constraint C9 can be satisfied by using the outputs of the neural network as the phase shift values. 
Problem (\ref{safe_rl_pro}) can be transformed into a convex problem by applying quadratic transform \cite{shen2018fractional} to the common and private rate expressions.
In the proposed RavNet, we solve problem (\ref{safe_rl_pro}) using the DCO layer \cite{agrawal2019}.
Compared with conventional convex solvers (e.g., CVX), the DCO layer 
can be integrated as a layer in RavNet.
In addition, it can solve problem (\ref{safe_rl_pro}) efficiently in a batch-wise manner, which significantly facilitates the training process.
We denote the feasible beamforming and IRS phase shift solutions obtained by solving problem (\ref{safe_rl_pro}) as $\B(\tau)$ and $\psmv(\tau)$, respectively.

We then feed $\B(\tau)$ and $\psmv(\tau)$ into a second DCO layer which solves the following optimization problem to obtain the RS parameters and bitrate selections:
\begin{equation}\label{jrsbr_pro}
\begin{aligned}
\underset{\substack{\br_n,\,n\in\N,\\c_i,\,i\in\I,\\p_{n,i},\,i\in\I_n,\,n\in\N}}{\text{maximize}} \quad &  \sum_{n\in\N} \left(\sum_{i\in\I_n} v_{n,i} - \kappa^{\text{intra}} \ell^{\text{intra}}_{n}\right) \\
\text{subject to } \quad & \text{constraints C1, C3$-$C8}.
\end{aligned}
\end{equation}
Note that the intra-frame quality switch loss $\ell^{\text{intra}}_{n}$ is a convex function with respect to the bitrate selection variables in vector $\br_n$.
Moreover, $R^{\ctxt}(\tau)$ and $R_n^{\ptxt}(\tau),n\in\N,$ in problem (\ref{jrsbr_pro}) can be determined given $\B(\tau)$ and $\psmv(\tau)$. 
All constraints except constraint C1 are affine constraints.
We relax constraint C1 as 
\begin{equation}
v_1\leq \, v_{n,i}\leq v_M,\,i\in\I_n,\,n\in\N.
\end{equation}
The relaxed problem is a convex optimization problem and can be solved using the DCO layer.
We round down the solution of $v_{n,i},i\in\I_n,\,n\in\N,$ to the nearest feasible solution.

\subsection{Critic Network Structure}
The proposed critic network has a similar structure as the actor network.
Since the critic network approximates the Q-value, the layers for obtaining the feasible actions in the actor network are not required in the critic network.
This leads to the following two modifications: (a) the DCO layers are not present in the critic network, and (b) the tanh activation layer in the MLP module is replaced by the ReLU activation layer to generate the Q-value.
Although all the $V$ critic networks have the same network structure, their initial learnable parameters are different.
By doing this, we can obtain $V$ independent approximations of the state-action value function.
While it is a known problem that the state-action value function approximated by each critic network may suffer from overestimation, this problem can be tackled by using multiple (i.e., $V>1$) critic networks to obtain multiple estimates.
The minimum among these estimates is then used as the target for updating the learnable parameters.
We note that this technique has been applied in several state-of-the-art DRL algorithms, including \cite{TD32018scott,Tuomas2018soft,Ling2020soft,Meng2021effect}, to address the overestimation of the state-action value function.
We denote the hidden dimensionality of the MLP module in the critic network as $d_{\text{crt}}$.
% \textcolor{blue}{
% We denote the sizes of the three FC layers in the critic network as $d_{\text{crt1}}\times  d_{\text{crt2}}$, $d_{\text{crt2}} \times d_{\text{crt3}}$, and $d_{\text{crt3}}\times d_{\text{crt\_out}}$, respectively.
% }

\subsection{Computational Complexity}
\label{section:complexity}
The computational complexity for the policy learning in the proposed Deep-GRAIL algorithm depends on the following processes: (a) obtaining the demonstration replay for imitation learning, (b) updating the policy learned by the actor network, i.e., $\pi_{\TA_{\text{act}}}$, and (c) updating the state-action value function approximated by each of the $V$ critic networks, i.e., $Q_{\TA^{(m)}_{\text{crt}}},\,m=1,\ldots,V$.

\subsubsection{Imitation learning and demonstration replay}
In order to obtain the demonstration replay for imitation learning, the AO algorithm presented in the Appendix needs to be executed for $D \tau^{\text{max}}$ iterations.
Based on the analysis of the computational complexity of the AO algorithm, see (\ref{AO_complexity}) in the Appendix, the imitation learning process incurs the following computational complexity \cite[Section 1.3]{boyd2004convex}:
\begin{align}\label{complexity_imi}
\mathcal{O}_{\textrm{IMI}} &= \mathcal{O}\left(D \tau^{\text{max}}(C_{\textrm{BF}} N_t^3 N^3 + C_{\textrm{PS}} L^{4.5} \log(1/\epsilon) + N^3)\right)\notag\\
&= \mathcal{O}\left(D \tau^{\text{max}}(C_{\textrm{BF}} N_t^3 N^3 + C_{\textrm{PS}} L^{4.5} \log(1/\epsilon))\right),
\end{align}
where $C_{\textrm{BF}}$ and $C_{\textrm{PS}}$ denote the number of iterations of the FP-based beamforming algorithm and the FP-based phase shift control algorithm invoked in each iteration of the AO algorithm, respectively. In addition, $\epsilon$ is a positive constant denoting the solution accuracy \cite{luo2010semidefinite}.

\subsubsection{Actor-Critic Method and DDPG for Policy Learning} In each training iteration, updating the policy learned by the actor network, i.e., $\pi_{\TA_{\text{act}}}$, incurs a computational complexity of
\begin{equation}
\begin{aligned}
\mathcal{O}_{\textrm{act}} = \;\;\mathcal{O}\Big(&k^2_1 ch_1 +  k^2_2 ch_1 ch_2 + k^2_3 ch_2 ch_3 +N k^2_4 ch_4 \\
&+   k^2_5 ch_4 ch_5 + k^2_6 ch_5 ch_6 + d^2_{\text{act}} + N_t N + L \Big).
\end{aligned}
\end{equation}

Updating the state-action value function approximated by each of the $V$ critic networks, i.e., $Q_{\TA^{(m)}_{\text{crt}}},\,m=1,\ldots,V$, incurs the following computational complexity:
\begin{equation}
\begin{aligned}
\mathcal{O}_{\textrm{crt}} = \;\;\mathcal{O}\Big(&k^2_1 ch_1 +  k^2_2 ch_1 ch_2 + k^2_3 ch_2 ch_3  +N k^2_4 ch_4 \\
&+   k^2_5 ch_4 ch_5 + k^2_6 ch_5 ch_6 + d^2_{\text{crt}} \Big).
\end{aligned}
\end{equation}

Given the total number of training iterations $T^{\text{max}} \tau^{\text{max}}$, the overall computational complexity for the policy learning in the proposed Deep-GRAIL algorithm is:
\begin{equation}
\begin{aligned}\label{deepcomplex}
\mathcal{O}\Big(&D \tau^{\text{max}}(C_{\textrm{BF}} N_t^3 N^3 + C_{\textrm{PS}} L^{4.5} \log(1/\epsilon)) + T^{\text{max}} \tau^{\text{max}} V \\
&(k^2_1 ch_1   +k^2_2 ch_1 ch_2 + k^2_3 ch_2 ch_3 +N k^2_4 ch_4 +   k^2_5 ch_4 ch_5 \\
&+ k^2_6 ch_5 ch_6 + d^2_{\text{crt}}) + T^{\text{max}} \tau^{\text{max}} (d^2_{\text{act}} + N_t N + L)\Big).
\end{aligned}
\end{equation}

Our analysis shows that the number of reflecting elements $L$ has a more significant impact on the computational complexity of imitation learning than the number of antennas $N_t$ and the number of users $N$.
Apart from the aforementioned variables, the computational complexity of the actor-critic method and DDPG also increases with the dimensionality of the DNNs.
In addition, while training $V$ critic networks can mitigate the overestimation issue, we observe from (\ref{deepcomplex}) that a larger $V$ also leads to a higher computational complexity.

\textit{Remark 5:} In order to construct the common and private messages, the base station needs to determine the values of the RS parameters, i.e., $c_i,\,i\in\I$ and $p_{n,i},\,i\in\I_n,\,n\in\N$, by solving problem (\ref{jrsbr_pro}).
Compared to the systems without RS, this process incurs an additional computational complexity of $\mathcal{O}\left(N^3\right)$.

\color{black}

\section{Performance Evaluation}
We consider a 10 m $\times$ 10 m $\times$ 3.5 m indoor facility for VR streaming as illustrated in Fig. \ref{scenario_just}.
Each user is designated a 2.7 m $\times $ 2.7 m area  \cite{oculuslink_art}.
The base station is installed at the center of the ceiling, and the IRS is installed on one side of the wall at the midpoint between the ceiling and the floor.
%Our simulation setting is illustrated in Fig. \ref{VR_setup}.
We assume all channels, including the channels between the base station and the users, are line-of-sight (LoS) based on the aforementioned deployments of the base station and the IRS.
%While some existing works considered the absence of LoS channels due to blockages, 
We consider the presence of LoS channels in our simulations to investigate the full potential of the proposed IRS-aided RS VR system. 
We assume a carrier frequency of 60 GHz as this value is used in several commercial wireless VR systems, see, e.g., \cite{steamvr,vivevr}.
%A 60 GHz channel environment \cite{chong2015multii,ju2021mmw} is considered in our simulation to offer a high downlink channel capacity for VR streaming.
Let $d_{n,D}$, $d_{n,R}$, and $d_{0}$ denote the distance between the base station and user $n$, the distance between the IRS and user $n$, and the distance between the base station and the IRS, respectively.
We determine the CSI of the direct and reflected channels by $\h_{n,D} = (\frac{\nu}{4\pi d_{n,D}})^\zeta \widehat{\h}_{n,D}$, $\h_{n,R} = (\frac{\nu}{4\pi d_{n,R}})^\zeta \widehat{\h}_{n,R}$, and $\Gm = (\frac{\nu}{4\pi d_{0}})^\zeta \widehat{\Gm}$, where $\nu$ is the wavelength of the carrier signal and $\zeta$ is the pathloss exponent.
The elements in $\widehat{\h}_{n,D}$, $\widehat{\h}_{n,R}$, and $\widehat{\Gm}$ are complex Gaussian distributed with zero mean and unit variance.
The other simulation parameter settings are given in Table \ref{simupara}.
%\begin{figure}[!t]
%\centering
%\includegraphics[width=3.0in]{VR_setup.pdf}
%%\vspace{-5mm}
%\caption{Simulation setting for the IRS-aided rate-splitting VR streaming system. We assume the base station is connected with the VR server and the IRS via wired connections, which allow video data and control signaling transmissions.
%}
%\label{VR_setup}
%\end{figure}

\begin{figure} 
    \centering
  \subfloat[360-degree video streaming session 1 \label{1a}]{%
       \includegraphics[width= 2.6in]{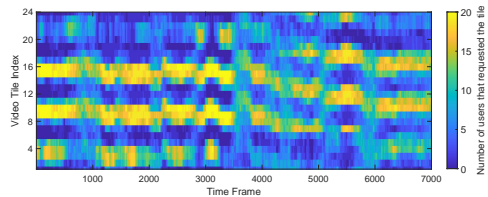}}
    \hfill
  \subfloat[360-degree video streaming session 2 \label{1b}]{%
        \includegraphics[width= 2.6in]{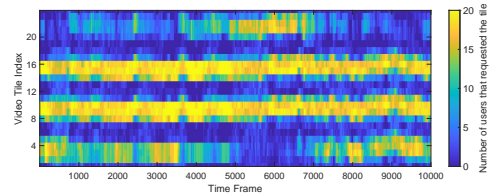}}
  \caption{Visualization of the video tile requests obtained from two VR streaming sessions in the real-world dataset \cite{knorr2018vr}. Each video frame is divided into $24$ video tiles. The $x$-axis indicates the video frame, while the time duration of each frame is $1$ sec. The $y$-axis shows the indices of the video tiles (i.e., from $1$ to $24$). The color of the grid centered at $(x,y)$ indicates the number of users that requested the $y$-th tile in the $x$-th video frame.}
  \label{interests_showcase} 
\end{figure}

To properly model the pattern of the video tiles requested by the users during the VR streaming session, we use the real-world dataset from \cite{knorr2018vr} to determine the video tile requests in our simulations.
The dataset from \cite{knorr2018vr} includes the head movements of $20$ users during multiple real-world VR streaming sessions.
The head movement record of a particular user is used to determine the FoV and the video tile requested by this user.
We divide each 360-degree video frame into $24$ tiles, with $N_x=4$ and $N_y =6$.
The FoV of each user covers $110$ degrees in horizontal direction and $90$ degrees in vertical direction of the video frame.
In Fig. \ref{interests_showcase}, we visualize the video tile requests that we determined based on two VR streaming sessions from the real-world dataset \cite{knorr2018vr}.
We use FFmpeg \cite{ffmpeg} to encode the 360-degree video into different bitrates as given by set $\V=\{2,3,4,5,6,7,8,9,10,11\}$ Mbps.
\begin{table}[t!]
  \begin{center}
    \caption{Simulation Parameters for Performance Evaluation}
    \begin{tabular}{|C{52mm}|C{24mm}|}
  	  \hline
      Parameter & Value\\
      \hline
      Bandwidth for downlink $W$ & $1$ GHz\\
      \hline
      Path loss exponent $\zeta$ & $2.29$ \cite{ju2021mmw} \\ 
      \hline
      Maximum transmit power $P^{\text{max}}$ & $1$ Watt\\
      \hline
      Noise power & $-174$ dBm/Hz\\
      \hline
      Time duration of each video tile $T_v$ &   $1$ sec \\
      \hline
      Downlink transmission window $\TD$ & $10$ ms\\
      \hline
      Coefficient for inter-frame quality switch loss $\kappa^{\text{intra}}$ & $10$ \\
      \hline
      Number of decision epochs per time slot (i.e., per episode) $\tau^{\text{max}}$ & $50$\\
      \hline
      Learning rate $\alpha$ & $5\times 10^{-4}$\\
      \hline
      Minibatch size $M_D$ & $512$\\
      \hline
      Number of critic networks $V$ & $6$\\
      \hline
      Value of $q$ for $q$-step return & $5$\\
      \hline
      Kernel size of the CNN layers & $2 \times 2$ \\
      \hline
      Number of channels of the CNN layers $ch_1$, $ch_2$, $ch_3$, $ch_4$, $ch_5$, $ch_6$  & $16$, $16$, $32$, $16$, $16$, $32$  \\
      \hline
      
      Hidden dimensionality of the MLP modules $d_{\text{act}}$,  $d_{\text{crt}}$ & $1024$ \\
      \hline
      Coefficients for training loss $\omega_1$, $\omega_2$ & $10^{-3}$ , $1$\\
      \hline
      Coefficient for soft update $\kappa$ & $5\times 10^{-3}$ \\
      \hline
      Discount factor $\gamma$ & $0.95$ \\
      \hline
    \end{tabular}
    \label{simupara}
  \end{center}
\end{table}

We conduct the simulation using a computing server with an Intel Core i5-9500 @ 3.0 GHz CPU, and an NVIDIA GeForce RTX 2070 GPU with 8 GB memory.
The codes for the proposed Deep-GRAIL algorithm are available on  \url{https://github.com/ruihuang1967/Deep-GRAIL}.

We compare the performance of the following baseline systems and algorithms:
\begin{itemize}
%\item The proposed C-DRL algorithm.

\item \textbf{IRS-aided RS VR streaming system with AO algorithm}: We use the AO algorithm presented in the Appendix to solve problem (\ref{qoe_pro}).

\item \textbf{IRS-aided RS VR streaming system with supervised learning (SL) algorithm}: In this algorithm, we train a DNN module using SL to minimize the mean squared error between its output and the solution of the AO algorithm presented in the Appendix.
This DNN module uses the same network structure as the proposed actor network.

\item \textbf{IRS-aided RS-NOUM system \cite{mao2019rate}}: We extend the RS-NOUM system proposed in \cite{mao2019rate} by including an IRS.
In this system, the information of the requested tiles are sent to all users via multicast. 
Each user also receives a dedicated unicast message regarding its requested tiles.
The multicast and unicast messages are combined using RS.
We solve the sum-rate maximization problem for the resulting system with the constraints of problem (\ref{qoe_pro}) using an AO-based algorithm with weighted minimum mean square error (WMMSE), SDR, and convex optimization.

\item \textbf{IRS-aided multiuser system without RS (IRS-aided MU system) \cite{wu2019irs}}: In this system, the requested video tiles are sent to the users via unicast without RS.
We solve the sum-rate maximization problem for the resulting system with the constraints of problem (\ref{qoe_pro}) using an AO-based algorithm with FP, SDR, and convex optimization.

\end{itemize}

\subsection{Convergence of the Deep-GRAIL Algorithm}
\begin{figure}[!t]
\centering
\includegraphics[width=3.0in]{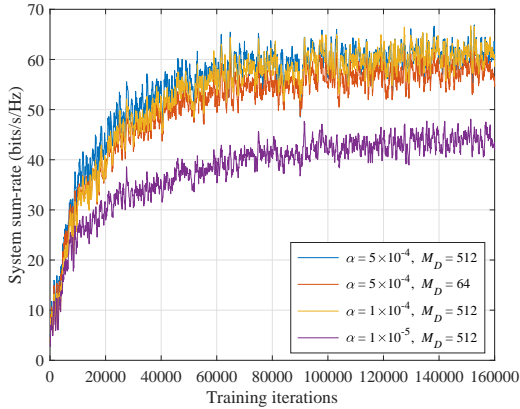}
\caption{ Convergence of the proposed Deep-GRAIL algorithm. We set $N_t=6$, $N=6$, and $L=100$.}
\label{convergence}
\end{figure}

We first investigate the convergence of the proposed Deep-GRAIL algorithm.
We show the achievable system sum-rate of the Deep-GRAIL algorithm versus the number of training iterations in Fig. \ref{convergence}.
We observe that with a properly chosen learning rate $\alpha$, e.g., $5 \times 10^{-4}$, the proposed Deep-GRAIL algorithm can efficiently improve the learned policy.
The results in Fig. \ref{convergence} also show that setting $\alpha$ to be too small, e.g., $10^{-5}$, can lead to slow convergence and inefficient policy learning.
In addition, we observe that increasing the minibatch size $M_D$ from $64$ to $512$ leads to a higher system sum-rate.

\subsection{Achievable System Sum-Rate}
\begin{figure}[!t]
\centering
\includegraphics[width=3.0in]{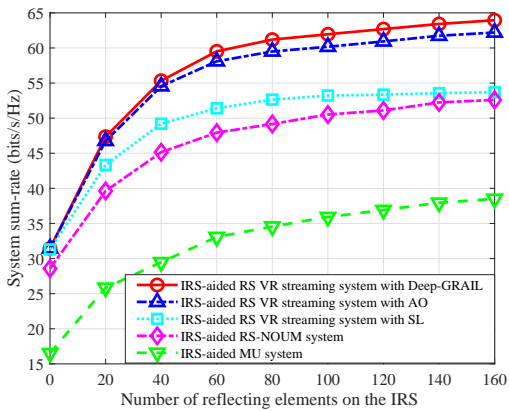}
\caption{System sum-rate versus the number of reflecting elements $L$. We set $N_t=N =6$. Note that $L=0$ represents the system without an IRS.}
\label{varyingL}
\end{figure}
In Fig. \ref{varyingL}, we vary the number of reflecting elements $L$ and investigate the system sum-rate.
We observe that the performance improvement of the proposed IRS-aided RS VR streaming system with Deep-GRAIL algorithm over the same system with AO algorithm increases with the value of $L$.
This is because when the IRS phase shift subproblem is solved using FP and SDR, Gaussian randomization is needed to obtain IRS phase shift matrix, which may incur significant performance degradation \cite{luo2010semidefinite}.
The proposed Deep-GRAIL algorithm avoids such performance loss since (a) SDR  is not required for the Deep-GRAIL algorithm, and (b) the imitation loss in (\ref{imitation_loss}) prevents the learning agent from being affected by the suboptimality of the AO algorithm with SDR.
In particular, when $L=160$, the IRS-aided RS VR streaming system with the proposed Deep-GRAIL algorithm achieves a system sum-rate that is $2.8\%$, $19.1\%$, $21.6\%$, and $66.1\%$ higher than that of the IRS-aided RS VR streaming system with AO algorithm, IRS-aided RS VR streaming system with SL algorithm, IRS-aided RS-NOUM system, and IRS-aided MU system, respectively.

In addition, in Fig. \ref{varyingL}, $L=0$ implies a system without IRS.
We observe that all considered systems benefit significantly from having an IRS present for improving the system sum-rate.
For the IRS-aided RS VR streaming system with Deep-GRAIL algorithm, deploying an IRS with $L=100$ reflecting elements results in a system sum-rate improvement of $91.44\%$ compared to the same system without IRS.
This is due to the SINR improvement achieved with the additional propagation channels created by the IRS.
Furthermore, for the proposed IRS-aided RS VR streaming system, since the common rate is determined by the user experiencing the minimum SINR (as shown in (\ref{common_rate_c})), the additional DoF introduced by the IRS are implicitly exploited to increase the rate of the common message.
In particular, our results show that the average of the achievable rate of the common message of the users, i.e., $R^{\ctxt}$ in (\ref{common_rate_c}), in the proposed IRS-aided RS VR system with $L=100$ reflecting elements is $12.81$ bits/s/Hz, while only an average of $5.61$ bits/s/Hz is achieved in the same system without IRS.
Therefore, using an IRS with $L=100$ reflecting elements increases the common rate by $128.3\%$, which allows more data to be transmitted via the common message to exploit the shared interests and improve the QoS of the users.
Our results demonstrate the benefits of IRS for mitigating the performance bottleneck of RS caused by the user experiencing the minimum SINR.

\begin{figure}[!t]
\centering
\includegraphics[width=3.0in]{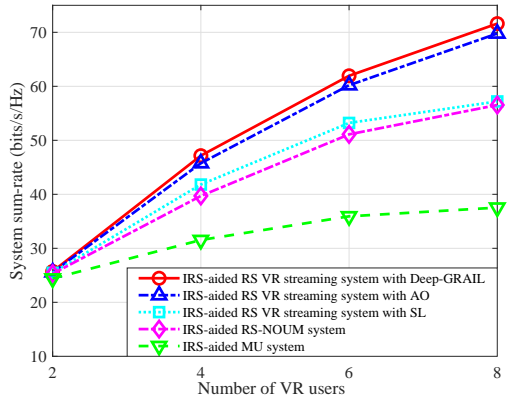} 
\caption{System sum-rate versus the number of VR users $N$. We set $N_t=6$ and $L = 100$.}
\label{AO_varyingN_plot}
\end{figure}

In Fig. \ref{AO_varyingN_plot}, we show the system sum-rate versus the number of users $N$. 
We set $N_t = 6$ and $L=100$. 
We observe that the performance gains of the IRS-aided RS VR streaming system over the IRS-aided RS-NOUM and IRS-aided MU systems become more pronounced with more users.
With more users, a particular tile is more likely to be requested by multiple users, and therefore there are more shared tile requests of the users to be exploited by the IRS-aided RS VR streaming system.
When $N=8$, the IRS-aided RS VR streaming system with the proposed Deep-GRAIL algorithm achieves a system sum-rate that is $2.6\%$, $25.2\%$,  $26.7\%$, and $90.8\%$ higher than that of the IRS-aided RS VR streaming system with AO algorithm, IRS-aided RS VR streaming system with SL algorithm, IRS-aided RS-NOUM system, and IRS-aided MU system, respectively.

\begin{figure}[!t]
\centering
\includegraphics[width=3.0in]{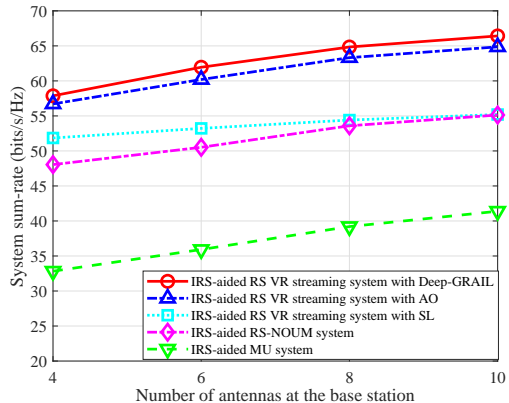}
\caption{System sum-rate versus the number of antennas $N_t$ at the base station. We set $N=6$ and $L=100$. }
\label{AO_varyingNt_plot}
\end{figure}

In Fig. \ref{AO_varyingNt_plot}, we show the system sum-rate of the users versus the number of antennas $N_t$ at the base station.
All algorithms except the IRS-aided RS VR streaming system with SL algorithm exhibit a similar performance gain as the number of antennas increases. 
The IRS-aided RS VR streaming system with SL algorithm suffers a larger sum-rate degradation when $N_t$ becomes larger.
This is caused by the increase in the mean squared error between the beamforming vectors determined by the SL algorithm and the beamforming vectors obtained by the AO algorithm.
When $N_t=10$, the IRS-aided RS VR streaming system with the proposed Deep-GRAIL algorithm achieves a $2.3\%$, $20.3\%$, $20.5\%$, and $60.5\%$ higher system sum-rate than the IRS-aided RS VR streaming system with AO algorithm, IRS-aided RS VR streaming system with SL algorithm, IRS-aided RS-NOUM system, and IRS-aided MU system, respectively.
Compared with the baseline schemes, the performance gain of the IRS-aided RS VR streaming system is due to the optimization of the RS parameters, i.e., $c_i,i\in\I$, given the video tile requests of the users.
Through the optimization of $c_i$, the base station can properly determine the video tiles that should be included in the common message, as well as the proportion of the common message allocated to them, such that the utility is maximized.
\begin{table*}[t!]
  \begin{center}
    \caption{Average Runtime Comparison for Different Schemes}
\begin{tabular}{| C{7.7cm}  | C{2.0cm} | C{2.0cm} | C{2.0cm} | C{2.2cm} | C{2.2cm} |}
\hline
Parameter Settings      & $N_t=6$, $N=4$, $L=100$ &  $N_t=6$, $N=6$, $L=100$ & $N_t=6$, $N=6$, $L=160$ & $N_t=10$, $N=6$, $L=100$\\
\hline
IRS-aided RS VR streaming system with Deep-GRAIL algorithm  & 8.52 sec  & 10.38 sec &  13.71 sec &  8.65 sec\\
\hline
IRS-aided RS VR streaming system with SL algorithm  & 0.75 sec  & 1.15 sec & 1.64 sec &  0.85 sec\\
\hline
IRS-aided RS VR streaming system with AO algorithm & 14.01 min   & 16.36 min & 63.32 min &  19.46 min\\
\hline
IRS-aided RS-NOUM system  &  12.98 min & 14.93 min & 59.49 min &  18.24 min \\
\hline
IRS-aided MU system  &  9.36 min & 10.20 min & 41.70 min &  13.21 min \\
\hline
\end{tabular}
\label{runtime_compare}
  \end{center}
\end{table*}
\subsection{Bitrate Allocation per User}
\begin{figure}[!t]
\vspace{4mm}
\centering
\includegraphics[width=2.65in]{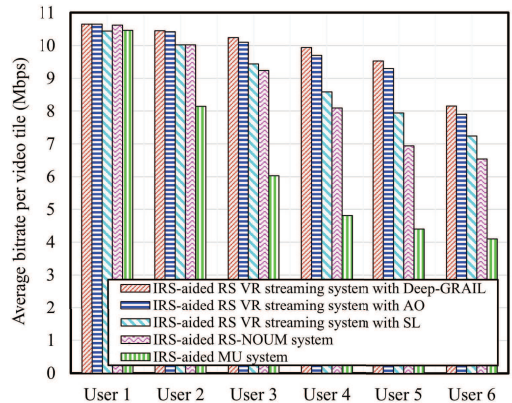}
\caption{Average achievable bitrate for each user. We set $N_t=N =6$ and $L=100$.}
\label{AO_UE_rate}
\end{figure}

In Fig. \ref{AO_UE_rate}, we show the average bitrate per video tile for each user.
We set $N_t=N =6$ and $L=100$.
We sort the users in descending order of their average bitrates.
That is, the user with the highest average bitrate is referred to as user $1$, while the user with the lowest average bitrate is referred to as user $6$.
The results in Fig. \ref{AO_UE_rate} show that the users achieve higher bitrates in the IRS-aided RS-enabled systems compared to the IRS-aided MU system.
This is because, with RS, the common message can be exploited to improve the QoS of multiple users simultaneously when those users have shared video tile requests.

\begin{figure}[!t]
\vspace{2mm}
\centering
\includegraphics[width=2.7in]{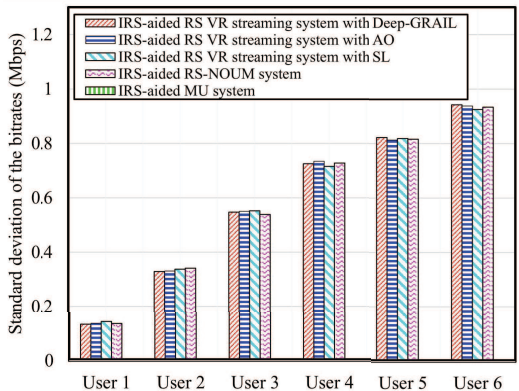}
\caption{Standard deviation of the bitrates for the video tiles. We set $N_t=N=6$ and $L=100$. The standard deviation for the IRS-aided MU system is zero for all users because solving the bitrate selection subproblem for this system yields the same bitrates for all tiles requested by a particular user.}
\label{UE_Var}
\end{figure}

In Fig. \ref{UE_Var}, we show the standard deviation of the bitrates for the video tiles received by the users.
For the IRS-aided MU system, solving the bitrate selection subproblem yields the same bitrates for all video tiles requested by a particular user.
This is due to the fact that (a) the IRS-aided MU system does not employ RS-based video tile transmission, and (b) the consideration of the intra-frame quality switch loss in the objective function $u(t)$ encourages the base station to minimize the standard deviation of the bitrates of the video tiles. 
Therefore, as can be observed in Fig. \ref{UE_Var}, the standard deviation for the IRS-aided MU system is zero for all users.
For the RS-enabled systems, those users with lower average bitrates (e.g., users 5 and 6) experience higher standard deviations of the bitrates of the received video tiles.
This is because by exploiting the common message, those users can obtain higher bitrates for video tiles that are requested by multiple users than for video tiles that are requested only by an individual user.
\subsection{Runtime Comparison}
In Table \ref{runtime_compare}, we compare the online execution runtime of different algorithms per time slot.
We observe that the average runtimes of the learning-based algorithms, i.e., the Deep-GRAIL and SL algorithms, are lower than that of the AO algorithms.
Moreover, the increases in runtime with respect to the value of $N_t$, $N$, and $L$ for the learning-based algorithms are less significant than the AO algorithms.
This is because the computationally expensive processes needed for solving the beamforming and IRS phase shift subproblems using WMMSE and SDR are not needed in the learning-based algorithms.
In particular, when $N_t=6$, $N=6$, and $L=160$, the average runtime of the Deep-GRAIL algorithm is only $0.36\%$, $0.38\%$, and $0.55\%$ of the average runtimes of the IRS-aided RS VR streaming system with AO algorithm, IRS-aided RS-NOUM system, and IRS-aided MU system, respectively.
The average runtime of the SL algorithm is lower than that of the Deep-GRAIL algorithm since the Deep-GRAIL algorithm needs to be executed for $\tau^{\text{max}}$ decision epochs per time slot, while the SL algorithm is not an iterative algorithm and only needs to be executed once per time slot.

\subsection{Ablation Experiment}
\begin{figure}[!t]
\centering
\includegraphics[width=3.0in]{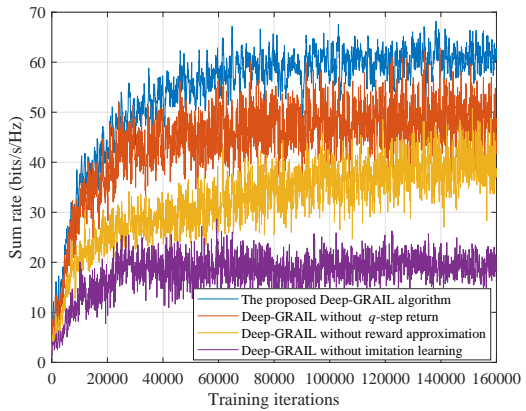} 
\caption{Ablation experiment for evaluating the effectiveness of the reward approximation, $q$-step return, and imitation learning in the proposed Deep-GRAIL algorithm.}
\label{ablation_test}
\end{figure}

We conduct an ablation experiment to investigate the effectiveness of the following components of the proposed Deep-GRAIL algorithm: (a) the reward approximation in (\ref{qoe_c_trans}), (b) the $q$-step return in (\ref{tar_Q_value}), and (c) the imitation loss in (\ref{imitation_loss}).
The results for the ablation experiment are shown in Fig. \ref{ablation_test}, where we compare the performance obtained after removing each of the three components from the proposed Deep-GRAIL algorithm.
We first observe that imitation learning offers the highest performance improvement among the three components. 
Without imitation learning, the agent learns a suboptimal policy, based on which only a sum-rate of approximately $20$ bits/s/Hz can be achieved.
This demonstrates that imitation learning can help the learning agent explore the state and action spaces more efficiently, and therefore discover a better policy.
We also observe that both the $q$-step return and reward approximation contribute to the improvement of the system sum-rate.
In particular, when reward approximation is not used, we observe a lower rate of convergence of the policy learning.
This is because without reward approximation, the reward function can be sparse due to the discrete bitrate selections.
Such sparsity can affect the learning efficiency.
The results in Fig. \ref{ablation_test} also show the benefits of using the $q$-step return to mitigate the overestimation of the state-action value function.
Without the $q$-step return, the actual total reward obtained by a chosen action can be lower than the one approximated by the state-action value function.
Since the policy learned by the actor network depends on the approximated state-action value function, such an inaccurate approximation can lead to performance degradation.

\subsection{Impact of Imperfect Channel Estimation}
\begin{figure}[!t]
\centering
\includegraphics[width=3.0in]{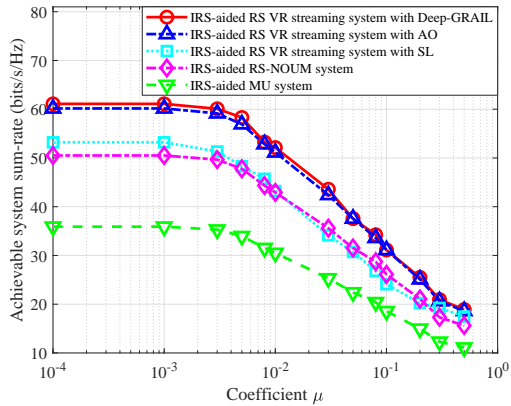}
\caption{Achievable system sum-rate for imperfect channel estimation. We set $N_t=N=6$ and $L=100$.}
\label{ICE_varying_mu}
\end{figure}

For the deviation of the proposed Deep-GRAIL algorithm, we assumed that the base station has the perfect CSI of the users.
In this subsection, we investigate the robustness of the proposed Deep-GRAIL algorithm to imperfect channel estimation using the statistical channel estimation error model from \cite{zhou2020framework,zhang2011multi}.
In particular, we use $\widehat{\F}(t) = \F(t) + \Delta \F(t)$, $\widehat{\h}_{n,D}(t) = \h_{n,D}(t) + \Delta \h_{n,D}(t)$, and $\widehat{\h}_{n,R}(t) = \h_{n,R}(t) + \Delta \h_{n,R}(t)$ to denote the estimated channel gain between the base station and the IRS, the estimated channel gain between user $n$ and the base station, and the estimated channel gain between user $n$ and the IRS, respectively.
For the channel estimation errors $\Delta \F(t)$, $\Delta \h_{n,D}(t)$, and $\Delta \h_{n,R}(t)$, the elements of $\Delta \F(t)$, $\Delta \h_{n,D}(t)$, and $\Delta \h_{n,R}(t)$ are assumed to follow complex Gaussian distributions with zero mean and variances $\mu^2 \norm{\mathrm{vec}(\F(t))}^2_2$, $\mu^2 \norm{\h_{n,D}(t)}^2_2$, and $\mu^2 \norm{\h_{n,R}(t)}^2_2$, respectively.
The coefficient $\mu\in[0,\,1)$ is a measure for the significance of the estimation error.
In Fig. \ref{ICE_varying_mu}, we evaluate the achievable system sum-rate for imperfect channel estimation.
We observe performance degradations in all considered algorithms due to imperfect channel estimation.
When $\mu$ is equal to $0.01$, the proposed Deep-GRAIL algorithm can retain $85.32\%$ of the system sum-rate that can be achieved for perfect channel estimation.
\color{black}

\section{Conclusion}
In this paper, we proposed a novel IRS-aided RS VR streaming system, in which the shared interests of the VR users were exploited via RS to improve the QoS of 360-degree video streaming.
We used IRS to create additional propagation channels, and improve the performance of RS by increasing the minimum SINR experienced by the common message at different users. 
We formulated the joint optimization of the RS parameters, IRS phase shifts, beamforming vectors, and bitrate selection as a mixed-integer nonlinear programming problem, in which the intra-frame quality switch loss and per-user per-tile QoS requirement were taken into consideration.
We proposed the Deep-GRAIL algorithm and RavNet, in which imitation learning, actor-critic method, DDPG, and DCO layers were employed to efficiently solve the formulated problem.
Simulation results based on a real-world dataset showed that the DoF introduced by RS and IRS can be efficiently exploited by the proposed Deep-GRAIL algorithm to achieve a higher system sum-rate compared to that of the IRS-aided RS-NOUM and IRS-aided MU systems.
The performance improvement of the proposed IRS-aided RS VR system became more pronounced in the presence of more shared video tile requests.
Our simulation results also revealed the respective contribution of RS and IRS to the performance gain achieved with the proposed IRS-aided RS VR system.
%In addition, we showed that the learning-based algorithms, i.e., the proposed Deep-GRAIL and the SL algorithms, required significantly less runtime than the existing AO algorithms. 
Through a runtime comparison with existing AO algorithms, we demonstrated the advantages of the proposed learning-based Deep-GRAIL algorithm in terms of runtime reduction, and its suitability for potential deployment in practical VR streaming systems.
For future work, we will extend the system model and solution approach to consider the IRS-aided RS VR streaming systems with imperfect CSI.
It is also interesting to take the subjective awareness of the users into account and study the quality of experience improvement achieved by IRS-aided RS VR streaming systems.

%\section*{Acknowledgment}
%This research was enabled in part by the support provided by the Digital Research Alliance of Canada (the Alliance) \cite{computecanada}.
%The authors also thank Ming Tang (tangm3@sustech.edu.cn) for her help and advice throughout the preparation of this paper.

\appendix
\section{The AO Algorithm for Solving Problem (\ref{qoe_pro})}
We decompose problem (\ref{qoe_pro}) into three subproblems where each of the subproblems can be solved by exploiting its hidden convexity.
For notational simplicity, we drop time index $t$ in this section.
We define $\boldsymbol{p} = (p_{n,i},\,i\in\I,n\in\N)$, $\boldsymbol{c} = (c_{i},\,i\in\I)$, and $\boldsymbol{v} = (v_{n,i},\,i\in\I,n\in\N)$.
While the objective function in problem (\ref{qoe_pro}) only depends on  bitrate selection $\br$, motivated by the inequality in (\ref{obj_trans}), we use the following function to take the effects of $\B$, $\psm$, $\boldsymbol{c}$, $\boldsymbol{p}$, and $\br$ on the objective function into consideration: 
\begin{align}
&r(\B,\psm,\boldsymbol{c},\boldsymbol{p},\br)\notag\\
& =\sum_{n\in\N} \frac{W \TD}{T_v} \Big(\sum_{i\in\I_n} p_{n,i} R_n^{\ptxt}(\B,\psm) + \sum_{i\in\I_n} c_i R^{\ctxt}(\B,\psm) \Big) \notag\\
& \hspace{5mm}- \sum_{n\in\N} \kappa^{\text{intra}}\,\ell_n^{\text{intra}}(\br)
\end{align}
For the beamforming subproblem, we optimize the beamforming vectors for maximization of $r(\B)$ subject to the maximum transmit power constraint C2 and the per-user per-tile QoS constraint C8.
We have
\begin{equation}\label{bf_pro}
\begin{aligned}
\underset{\substack{\B}}{\text{maximize}} \quad &  r(\B)\\
\text{subject to } \quad & \text{constraints C2 and C8}.
\end{aligned}
\end{equation}
Subproblem (\ref{bf_pro}) can be solved using WMMSE \cite{mao2021rate} or FP \cite{shen2018fractional} by introducing auxiliary variables, and updating the beamforming vectors and the auxiliary variables iteratively.
Both WMMSE and FP are guaranteed to converge to a stationary solution of problem (\ref{bf_pro}).
%In fact, for the beamforming subproblem (\ref{bf_pro}), the WMMSE algorithm and FP algorithm are equivalent \cite{shen2018fractional}.

%\subsection{Phase Shift Control Subproblem}
\newcommand{\psv}{\boldsymbol{\lambda}}
\newcommand{\mpsv}{\boldsymbol{\Lambda}}
\newcommand{\cjt}{\mathrm{H}}
\newcommand{\Hm}{\boldsymbol{\Theta}}
The IRS phase shift subproblem is given by 
\begin{equation}\label{ps_pro}
\begin{aligned}
\underset{\substack{\psm}}{\text{maximize}} \quad &  r(\psm)\\
\text{subject to } \quad & \text{constraints C8 and C9}.
\end{aligned}
\end{equation}
For the IRS phase shift subproblem, we define vector $\psv=(e^{-j\psi_{1}},\ldots,e^{-j\psi_{L}},\,\rho)\in\mathbb{C}^{L+1}$, where $\rho\in\mathbb{C}$ and $|\rho|^2=1$.
We further define matrix $\mpsv = \psv \psv^{H}\in\mathbb{C}^{(L + 1)\times (L+1)}$ to replace the IRS phase shift constraint C9.
This leads to the following equality constraints:
\begin{equation}\label{lamba_c0}
\text{C10:}\quad \mathrm{Diag}(\mpsv) = \boldsymbol{I}_{L+1},
\end{equation}
%\begin{equation}\label{lamba_c1}
%\mpsv \succeq \boldsymbol{0},
%\end{equation}
\begin{equation}\label{lamba_c2}
\text{C11:}\quad \rank(\mpsv) = 1,
\end{equation}
where $\boldsymbol{I}_{L+1}$ denotes the $(L+1) \times (L+1)$ identity matrix.
For user $n\in\N$, we define the following matrix
\begin{equation}
\Hm_n = \begin{bmatrix}
\diag(\h^{H}_{R,n})\,\F\\
\h^{H}_{D,n}
\end{bmatrix}\in \C^{(L+1)\times N_t}.
\end{equation}
To solve subproblem (\ref{ps_pro}) in a tractable manner, we rewrite the SINR of the common message of user $n$ in (\ref{common_sinr}) as follows:
\begin{equation}\label{common_sinr_trans}
\Gamma_{n}^{\ctxt} = \frac{\Tr(\mpsv^{T}\Hm_n \bfv_0 \bfv^{H}_0 \Hm^{H}_n)}{\sum_{j\in\N}\Tr(\mpsv^{T}\Hm_n \bfv_j \bfv^{H}_j \Hm^{H}_n) + \sigma^2}.
\end{equation}
The SINR of the private message of user $n$ in (\ref{pri_sinr}) can be rewritten  as
\begin{equation}\label{pri_sinr_trans}
\Gamma_{n}^{\ptxt} = \frac{\Tr(\mpsv^{T}\Hm_n \bfv_n \bfv^{H}_n \Hm^{H}_n)}{\sum_{j\in\N\setminus\{n\}}\Tr(\mpsv^{T}\Hm_n \bfv_j \bfv^{H}_j \Hm^{H}_n) + \sigma^2}.
\end{equation}
Similar to the beamforming subproblem, we use FP \cite{shen2018fractional} to tackle the multi-ratio fractional objective function in subproblem (\ref{ps_pro}). 
We apply quadratic transform \cite{shen2018fractional} to the common rate and the private rate of user $n\in\N$ as follows:
\begin{equation}
\begin{aligned}
\widetilde{R}^{\ctxt}_n = &\log_2\Big(1+2y_n\sqrt{\Tr(\mpsv^{T}\Hm_n \bfv_0 \bfv^{H}_0 \Hm^{H}_n)} \\
&- y_n^2\big(\sum_{j\in\N} \Tr(\mpsv^{T}\Hm_n \bfv_j \bfv^{H}_j \Hm^{H}_n) + \sigma^2\big)\Big),
\end{aligned}
\end{equation}
and
\begin{equation}
\begin{aligned}
\widetilde{R}^{\ptxt}_n = &\log_2\Big(1+2z_n\sqrt{\Tr(\mpsv^{T}\Hm_n \bfv_n \bfv^{H}_n \Hm^{H}_n)} \\
&- z_n^2\big(\sum_{j\in\N\setminus\{n\}} \Tr(\mpsv^{T}\Hm_n \bfv_j \bfv^{H}_j \Hm^{H}_n) + \sigma^2\big)\Big),
\end{aligned}
\end{equation}
where $y_n$ and $z_n$ are the auxiliary variables.
This leads to the following optimization problem 
\begin{align}\label{psmsubpro_fp}
\underset{\substack{\mpsv},\,\boldsymbol{y},\,\boldsymbol{z},\,\widetilde{R}^{\ctxt}}{\text{maximize}} \; & f^{\mathrm{PS}}(\mpsv,\widetilde{R}^{\ctxt},\boldsymbol{y},\boldsymbol{z})\notag\\
&\triangleq \sum_{n\in\N} \frac{W \TD}{T_v} \Big(\widetilde{R}^{\ptxt}_n + \sum_{i\in\I_n} c_i \widetilde{R}^{\ctxt}\Big) \notag\\
\text{subject to } \:\:\: 
& \widetilde{R}^{\ctxt} \geq 0\\
&\widetilde{R}^{\ctxt} \leq \widetilde{R}_n^{\ctxt},\,n\in\N\notag\\
& W \TD ( p_{n,i}  \widetilde{R}^{\ptxt} +  c_i \widetilde{R}^{\ctxt}) \geq T_v\, v_{n,i},\,i\in\I_n,\,n\in\N\notag\\
& \text{constraints C10 and C11}\notag,
\end{align}
\noindent \hspace{-1.5mm}where $\boldsymbol{y}=(y_1,\ldots,y_N)$ and $\boldsymbol{z}=(z_1,\ldots,z_N)$.
For user $n\in\N$, the optimal $y_n$ and $z_n$ for fixed $\mpsv$ are given by 
\begin{equation}\label{optimal_y}
y^\star_n = \frac{\sqrt{\Tr(\mpsv^{T}\Hm_n \bfv_0 \bfv^{H}_0 \Hm^{H}_n)}}{\sum_{j\in\N} \Tr(\mpsv^{T}\Hm_n \bfv_j \bfv^{H}_j \Hm^{H}_n) + \sigma^2},\,\,n\in\N,
\end{equation}
and
\begin{equation}\label{optimal_z}
z^\star_n = \frac{\sqrt{\Tr(\mpsv^{T}\Hm_n \bfv_n \bfv^{H}_n \Hm^{H}_n)}}{\sum_{j\in\N\setminus\{n\}}\Tr(\mpsv^{T}\Hm_n \bfv_j \bfv^{H}_j \Hm^{H}_n) + \sigma^2},\,\,n\in\N.
\end{equation}
For fixed $y_n$ and $z_n,\,n\in\N$, we use SDR \cite{luo2010semidefinite} to tackle constraint C11, and after relaxation the problem can be solved using convex optimization.
A suboptimal solution of subproblem (\ref{psmsubpro_fp}) can be obtained by iteratively optimizing $y_n,\,z_n,\,n\in\N$, and $\mpsv$ \cite{shen2018fractional}.
The FP-based algorithm for solving subproblem (\ref{psmsubpro_fp}) is provided in \textbf{Algorithm \ref{algo_psm}}.

%\subsection{Rate-Splitting and Bitrate Selection Subproblem}\label{sec_brs}
After $\psm$ and $\B$ have been determined, the RS parameters and bitrate selection can be obtained using the same approach as for solving problem (\ref{jrsbr_pro}).
We omit the details here for brevity.
We solve the aforementioned three subproblems iteratively until the objective function converges.
Since a feasible solution is required for the initialization of the AO algorithm, we propose the following method for obtaining a feasible solution.
We first initialize the bitrate selection to the minimum value.

That is,
\begin{equation}\label{v_ini}
v_{n,i}(t)=\begin{cases} v_1, \quad & i\in\I_n,\\
0, \quad &\text{otherwise}.
\end{cases}
\end{equation}
%We initialize the downlink transmission window $\TD$ to the maximum value, i.e., $\TD=\TD^{\text{max}}$.
We initialize vector $\boldsymbol{c}$ by splitting the common rate equally between the tiles in $\I$.
That is, 
\begin{equation}\label{c_ini}
c_i = \begin{cases} \frac{1}{|\I|}, \quad & i\in\I,\\
0, \quad &\text{otherwise}.
\end{cases}
\end{equation}
Similarly, we initialize $p_{n,i}$ as follows:
\begin{equation}\label{p_ini}
p_{n,i}(t) = \begin{cases} \frac{1}{|\I_n|}, \quad & i\in\I_n,\\
0, \quad &\text{otherwise}.
\end{cases}
\end{equation}

Then, we find a feasible beamforming solution by solving the following problem:
\begin{align}\label{bf_pro_ini}
\underset{\substack{\B}}{\text{maximize}} \quad &  r(\B) + \hspace{-1mm}\sum_{n\in\N} \hspace{-0.5mm}\sum_{i\in\I_n} \hspace{-1mm}  \eta \min\{R^{\ctxt} c_i + R_n^{\ptxt} p_{n,i} - v_{n,i}, 0\}\notag\\
\text{subject to } \quad & \text{constraint C2},
\end{align}
where $\eta>0$ is the scaling factor of the penalty from violating constraint C8.
Similar to the beamforming subproblem in (\ref{bf_pro}), problem (\ref{bf_pro_ini}) becomes a convex optimization problem after applying quadratic transform to $R^{\ctxt}$ and $R_n^{\ptxt}$.

After solving problem (\ref{bf_pro_ini}), we solve the following optimization problem to obtain a feasible IRS phase shift matrix:
\begin{align}\label{ps_pro_ini}
\underset{\substack{\psm}}{\text{maximize}} \quad &  r(\psm) +  \hspace{-1mm}\sum_{n\in\N} \hspace{-0.5mm}\sum_{i\in\I_n} \hspace{-1mm} \eta \min\{R^{\ctxt} c_i + R_n^{\ptxt} p_{n,i} - v_{n,i}, 0\}\notag\\
\text{subject to } \quad & \text{constraint C9}.
\end{align}
Problem (\ref{ps_pro_ini}) can be solved using FP and SDR.
The proposed iterative algorithm for obtaining an initial solution is shown in \textbf{Algorithm \ref{algo_ini}}.
We increase the value of $\eta$ by multiplying it with a scale factor $\beta>1$ after each iteration to increase the penalty for violating constraint C8. 
Typical values of $\beta$ range from $2$ to $10$.
We solve problems (\ref{bf_pro_ini}) and (\ref{ps_pro_ini}) iteratively until a feasible solution is found or the maximum number of iterations is reached.

\begin{algorithm}[!t]
%\algsetup{linenosize=\small}
\small
\caption{Algorithm for Phase Shift Subproblem (\ref{psmsubpro_fp})}\label{algo_psm}
\begin{algorithmic}[1]
\STATE Initialize $\psv$ to a feasible value $\psv^{(0)}$ and obtain $\mpsv^{(0)}$.
\STATE Initialize the FP termination threshold $\varepsilon_{\mathrm{FP}}$.
\STATE Initialize the iteration counter $\tau \leftarrow 0$.
\STATE Initialize $f^{\mathrm{PS}}(\mpsv^{(\tau)},(\widetilde{R}^{\ctxt})^{(\tau)},\boldsymbol{y},\boldsymbol{z}) \leftarrow 0$.
\REPEAT
\STATE Determine the values of $y^\star_n$, $z^\star_n,\,n\in\N$ based on (\ref{optimal_y}) and (\ref{optimal_z}), respectively.
\STATE $y_n \leftarrow y^\star_n, z_n \leftarrow z^\star_n,\,n\in\N$.
\STATE Solve subproblem (\ref{psmsubpro_fp}) for fixed $\boldsymbol{y}$ and $\boldsymbol{z}$, and obtain the optimal $\mpsv^{(\tau+1)}$ and $(\widetilde{R}^{\ctxt})^{(\tau+1)}$.
\STATE $\tau \leftarrow \tau + 1$.
\UNTIL $|f^{\mathrm{PS}}(\mpsv^{(\tau)},(\widetilde{R}^{\ctxt})^{(\tau)},\boldsymbol{y},\boldsymbol{z})$$-$$f^{\mathrm{PS}}(\mpsv^{(\tau-1)},(\widetilde{R}^{\ctxt})^{(\tau-1)},\boldsymbol{y},\boldsymbol{z})|$\\
$\leq \varepsilon_{\mathrm{FP}}$.
\STATE Decompose $\mpsv^{(\tau)}$ to obtain the phase shift matrix $\psm^{\star}$.
\end{algorithmic}
\end{algorithm}
\begin{algorithm}[!t]
%\algsetup{linenosize=\small}
\small
\caption{Algorithm for Obtaining an Initial Solution}\label{algo_ini}
\begin{algorithmic}[1]
\STATE Initialize $\br$ based on (\ref{v_ini}).
\STATE Initialize $\boldsymbol{c}$ and $\boldsymbol{p}$ based on (\ref{c_ini}) and (\ref{p_ini}), respectively.
\STATE Initialize $\B^{(0)}$ and $\psm^{(0)}$ based on random initialization.
\STATE Initialize $\eta \leftarrow \eta_0$.
\STATE Initialize iteration counter $\tau \leftarrow 1$.
\FOR{$\tau \leq \tau^{\text{max}}$}
\STATE Solve problem (\ref{bf_pro_ini}) and obtain solution $\B^{(\tau)}$.
\STATE Solve problem (\ref{ps_pro_ini}) and obtain solution $\psm^{(\tau)}$.
\IF{$(\B^{(\tau)},\psm^{(\tau)})$ is a feasible solution}
 \STATE break.
\ENDIF
\STATE $\tau \leftarrow \tau+1$.
\STATE $\eta \leftarrow \beta \eta$.
\ENDFOR
\end{algorithmic}
\end{algorithm}

For the computational complexity analysis of the AO algorithm, we assume that $\sum_{n\in\N} I_n$ and $|\I|$ increase linearly with respect to the number of users $N$.
For convex optimization, we assume that the computational cost of evaluating the first and second derivatives of the objective and constraint functions are dominated by the remainder of the convex optimization process  \cite[Section 1.3]{boyd2004convex}.
In each iteration of the FP-based beamforming algorithm, a convex optimization problem with $N_t(N+1)$ optimization variables and $\sum_{n\in\N} I_n + 1$ constraints is solved.
Using the interior-point method, this incurs a computational complexity of $\mathcal{O}_{\textrm{BF}}=\mathcal{O}(N_t^3 N^3)$ \cite[Section 1.3]{boyd2004convex}.
Moreover, an SDR problem is solved in each iteration of the IRS phase shift control algorithm.
The worst case complexity of solving such problem is $\mathcal{O}_{\textrm{PS}}=\mathcal{O}(L^{4.5} \log(1/\epsilon))$, where $\epsilon>0$ is a constant denoting the solution accuracy \cite{luo2010semidefinite}.
The joint RS parameter optimization and bitrate selection subproblem is a convex optimization problem with $2\sum_{n\in\N} I_n + |\I|$ variables and $3\sum_{n\in\N} I_n +|\I|+ N + 2$ constraints.
Solving this subproblem incurs a  computational complexity of $\mathcal{O}_{\textrm{RS}}=\mathcal{O}\left(N^3\right)$.

Let $C_{\textrm{BF}}$, $C_{\textrm{PS}}$, and $C_{\textrm{AO}}$ denote the number of iterations of the FP-based beamforming algorithm, the FP-based phase shift control algorithm, and the overall AO-based algorithm, respectively.
The overall computational complexity of the AO algorithm is 
\begin{align}\label{AO_complexity}
\mathcal{O}_{\textrm{AO}} &= \mathcal{O}\left(C_{\textrm{AO}}(C_{\textrm{BF}} N_t^3 N^3 + C_{\textrm{PS}} L^{4.5} \log(1/\epsilon) + N^3)\right)\notag\\
&= \mathcal{O}\left(C_{\textrm{AO}}(C_{\textrm{BF}} N_t^3 N^3 + C_{\textrm{PS}} L^{4.5} \log(1/\epsilon))\right).
\end{align}

\bibliographystyle{IEEEtran}
\bibliography{IEEEabrv,ruiURLLC,VR,ruiNCO,ruiGFMA,ruiJournal}

\end{document}